\documentclass[journal]{IEEEtran}
\usepackage{placeins}
\usepackage{diagbox}
\usepackage[inline]{enumitem}
\usepackage{xcolor}
\usepackage{amsmath,amssymb}
\usepackage{algorithm}
\usepackage[group-separator={,},group-minimum-digits=4]{siunitx}
\usepackage{threeparttable}
\usepackage{array}
\usepackage[caption=false,font=normalsize,labelfont=sf,textfont=sf]{subfig}
\usepackage{textcomp}
\usepackage{stfloats}
\usepackage{url}
\usepackage{verbatim}
\usepackage{graphicx}
\usepackage{algpseudocode}
\usepackage{cite}
\usepackage{balance}
\usepackage{mathrsfs}
\usepackage[colorlinks=true,
            linkcolor=blue,
            anchorcolor=blue,
            citecolor=blue]{hyperref}

\begin{document}

\title{ 
%\fontsize{16.5}{\baselineskip}\selectfont
{A Continuous Payload-Bearing Discrete Multitone Modulation Framework for Fiber-Optic Integrated Sensing and Communication} %With Spatial-ISI-Free Channel Reconstruction
} % \fontsize{20.5pt}{\baselineskip}\selectfont{xxx}
\author{ 
{%Authors
    Huan~Huang,~\textit{Member,~IEEE}, 
    Ziang~Chen,
    Zhiyang~Xue,
    Dongdong~Zou,
    % Zhongxing~Tian,
    %Gangxiang~Shen,~\textit{Senior~Member,~IEEE,~Fellow,~Optica}, 
    and Yi~Cai,~\textit{Fellow,~Optica}%\textit{Fellow,~Optica}
}% <-this % stops a space
\thanks{  
%This work was supported by the National Natural Science Foundation of China under Grant 62401387; in part by the Natural Science Foundation of Jiangsu Province under Grant BK20240768. (\textit{Corresponding author: Huan~Huang})

H.~Huang, Z.~Chen, Z.~Xue, D.~Zou, and Y.~Cai are with the School of Electronic and Information Engineering, Soochow University, Suzhou, Jiangsu 215006, China 
(e-mail: hhuang1799@gmail.com; zachen@stu.suda.edu.cn; ddzou@suda.edu.cn; zyxue999@stu.suda.edu.cn;  yicai@suda.edu.cn).   %Z.~Chen, D.~Zou, Z.~Tian,  zachen@stu.suda.edu.cn; ddzou@suda.edu.cn; zxtian@ieee.org;  shengx@suda.edu.cn; 
}
}
\maketitle

\begin{abstract}
We propose a continuous discrete multitone (DMT) framework for fiber-optic integrated sensing and communication (ISAC), in which the same payload-bearing waveform supports forward intensity-modulation/direct-detection (IM/DD) communication and backward distributed acoustic sensing without a separate sensing waveform or silent interval. A unified phase-sensitive optical time-domain reflectometry model represents distributed Rayleigh backscattering as a finite-memory sensing channel. It distinguishes the round-trip silent guard required to isolate pulsed returns from spatial intersymbol interference caused by off-peak matched-filter correlations. Cyclic-prefix DMT (CP-DMT) uses a cyclic prefix (CP) covering the sensing memory and one-tap frequency-domain equalization to eliminate this correlation-induced coupling under ideal full-bin inversion. Cyclic-prefix-free DMT (NoCP-DMT) reconstructs the long-memory linear convolution by regularized least squares, removing the CP at higher receiver complexity. %Both modes apply noise-whitened principal component combining before gauge-differential phase recovery.
Experiments over an approximately 10-km fiber link localize a 600-Hz disturbance applied by a piezoelectric transducer (PZT) near 5.071~km and recover gauge-differential phase with correlations of 0.9982 and 0.9983 for CP-DMT and NoCP-DMT, respectively. At 2- and 1-V PZT drive levels, the fifteen-record localization standard deviations are 0.43/0.41~m for CP-DMT and 0.38/0.38~m for NoCP-DMT. The corresponding IM/DD error vector magnitude values range from 4.79\% to 5.01\%, with no bit errors observed.
\end{abstract}


\section{Introduction}\label{Intro}
\IEEEPARstart{I}{ntegrated} sensing and communication (ISAC) aims to use a common physical infrastructure for both information delivery and environmental perception. Optical fiber is naturally suited to this objective because it combines high-capacity data transport over deployed networks~\cite{AgrawalBook2010,EssiambreJLT2010} with intrinsic sensitivity to strain, vibration, temperature, and acoustic disturbances~\cite{HartogBook2017}. In particular, phase-sensitive optical time-domain reflectometry (\(\phi\)-OTDR) with coherent detection enables distributed acoustic sensing (DAS) by recovering dynamic Rayleigh-backscatter responses along the fiber~\cite{LuJLT2010}, supporting applications such as distributed intrusion detection, industrial infrastructure monitoring, and seismic observation~\cite{JuarezJLT2005,AshryJLT2022,FernandezRuizJLT2022}. Deployed telecom cables can also be reused as wide-area distributed sensing media~\cite{IpJOCN2022}. Moving beyond infrastructure sharing, waveform-level fiber-optic ISAC seeks to make a common payload-bearing waveform serve both sensing and communication functions, consistent with the broader ISAC vision~\cite{LiuJSAC2022} and the dual-function-waveform paradigm in joint radar-communication systems~\cite{ZhangCOMST2022,HassanienTSP2016}.

At the sensing receiver, Rayleigh backscatter from distributed locations interferes coherently, causing the sensitivity to vary with position, while receiver noise further limits recovery of the distributed response~\cite{GabaiOL2016,VidalOE2023Noise,TLSMOTDR}. These effects complicate estimation of the complex distributed response required for subsequent acoustic phase extraction. At a stricter level of waveform sharing, the same continuous payload-bearing waveform is desired in order to support forward communication decoding and serve as the known excitation for backward distributed-channel reconstruction. Existing fiber-optic ISAC schemes can be broadly organized according to their dominant integration mechanism: resource multiplexing, communication-receiver reuse, or reuse of a structured waveform component. 

The first route is resource-multiplexed coexistence. Wang \emph{et al.} demonstrated co-propagation of real-time DAS and coherent 400-GbE traffic in adjacent dense wavelength-division-multiplexed channels of a field-deployed metro network~\cite{WangJLT2024}. Gu \emph{et al.} placed communication and sensing in the same wavelength channel but separated them by frequency-division multiplexing~\cite{GuPR2025}, whereas Wei \emph{et al.} combined space- and wavelength-division multiplexing in a field-deployed seven-core fiber~\cite{WeiJOCN2026}. These systems share the fiber infrastructure, but retain explicit separation in wavelength, frequency, or space.

The second route reuses the forward communication field and coherent-receiver digital signal processing (DSP) for sensing. Ip \emph{et al.} extracted vibration-induced phase using modified coherent transponders and localized disturbances through correlation~\cite{IpJLT2022Transponder}. In digital-subcarrier systems, Yang \emph{et al.} reused frequency-domain pilot tones for carrier recovery and forward phase sensing~\cite{YangJLT2025}; subsequent Wiener filtering improved vibration extraction under commercial external-cavity lasers~\cite{YangOL2025}. This route shares the payload-bearing optical field and receiver chain, but observes forward-field perturbations rather than reconstructing the backward distributed channel.

The third route reuses a deliberately structured component of the communication waveform as the sensing reference. He \emph{et al.} used a periodically chirped optical carrier to carry four-level pulse-amplitude modulation (PAM-4) data and simultaneously serve as the \(\phi\)-OTDR probe~\cite{HeLSA2023}. Hu \emph{et al.} repurposed a fractional-Fourier-transform training sequence, originally introduced for time and frequency synchronization, for distributed vibration sensing~\cite{HuOL2024}. This route therefore shares an explicitly designed carrier or training structure, rather than only infrastructure or receiver resources.

Together, these approaches extend fiber-optic ISAC beyond infrastructure coexistence toward receiver and waveform integration. However, they did not use payload symbols themselves as the known excitation for backward distributed-channel reconstruction. Martins \emph{et al.} addressed this gap by representing the coherent Rayleigh backscatter of phase-shift-keying (PSK) data as a linear convolution and estimating the fiber impulse response through least-squares (LS) channel estimation~\cite{MartinsOE2016}. Their 500-m proof of concept established feasibility, but applying the formulation to long sensing links yields a large, long-memory inverse problem. Its computational complexity scales with the channel memory, whereas the estimation conditioning depends on the finite observation sequence and the corresponding convolution matrix.

More recently, Liu \emph{et al.} demonstrated simultaneous forward communication and backward DAS using quadrature phase-shift keying (QPSK) streams in an optical supervisory channel~\cite{LiuOL2025}. The distributed response was recovered by cross-correlation, where zero padding longer than the round-trip time (RTT) was inserted to isolate successive data sequences. Consequently, the nonideal autocorrelation of the data sequence limits the range response, while the required padding reduces the payload duty cycle. A related combination of RTT-scale return isolation and correlation-based range recovery is also inherent to matched-filter (MF)-based pulse compression.

Specifically, the MF-based received trace is correlated with the transmitted probing waveform to obtain a range-resolved response~\cite{ZouOE2015,MompoOL2018}. A post-pulse silent interval no shorter than the fiber round-trip delay is therefore required to isolate successive returns.
However, off-peak samples of the pulse-compression kernel still couple different range bins and produce deterministic MF-correlation-induced spatial intersymbol interference (ISI), hereafter termed spatial ISI~\cite{MompoOL2018,HuangArxiv2026}.
A complementary approach is to simplify the long-memory channel inverse through cyclic block structure. A recent DAS scheme based on orthogonal frequency-division multiplexing (OFDM) with a cyclic prefix (CP) modeled backward Rayleigh scattering as a finite-memory sensing multipath channel~\cite{HuangArxiv2026}. When the CP covers the full sensing memory, the retained observation follows circular convolution, enabling one-tap frequency-domain equalization (FDE) and inverse-transform channel reconstruction. Because this approach does not retain the MF-based pulse-compression kernel, it avoids spatial ISI under sufficient-CP and ideal-inversion conditions.

However, the CP must cover the effective sensing-channel memory, which is dominated by the fiber RTT. Even a short-reach or access span may therefore have a sensing memory spanning many communication symbols, so repeating a full-memory CP on every multicarrier symbol introduces substantial redundancy. Extending the useful-symbol duration amortizes this overhead but increases latency and tightens the channel-stationarity requirement. 
In addition, the kilometer-scale operating range of DAS overlaps the reach of standardized intensity-modulation/direct-detection (IM/DD) optical access systems~\cite{ITUG98071}. This overlap motivates considering IM/DD access links as a practical forward-communication setting for waveform-level fiber-optic ISAC. Discrete multitone (DMT) modulation retains the multicarrier structure of OFDM~\cite{ShiehOFDMBook}, supports adaptive loading over spectrally shaped channels~\cite{BinghamCommMag1990,ChowTCOMM1995}, and has a mature standardization history in practical wireline systems~\cite{ITUG9921}. With Hermitian-symmetric loading, DMT yields a real-valued electrical waveform directly compatible with IM/DD optical links~\cite{ArmstrongJLT2009,LeeJLT2009}.

We develop a continuous payload-bearing DMT framework with two alternative modes for fiber-optic ISAC: cyclic-prefix DMT (CP-DMT) with full-memory-CP-assisted FDE and cyclic-prefix-free DMT (NoCP-DMT) with regularized long-memory channel reconstruction. Within either mode, the same transmitted DMT waveform carries the forward IM/DD payload and provides the known excitation for backward DAS.
The main contributions are summarized as follows:
\begin{itemize}
\item
We derive a unified \(\phi\)-OTDR model that maps the coherent Rayleigh return to a finite-memory sensing multipath channel whose taps represent physical range cells. The formulation separates physical channel memory from the waveform-dependent reconstruction operation and supports all three sensing receivers considered herein. It also distinguishes inter-pulse interference (IPI) caused by overlapping returns from MF-correlation-induced spatial ISI, which arises from nonzero off-peak correlation samples and cannot be removed by extending the silent guard interval.
\vspace{2pt}

\item
We develop continuous payload-bearing CP-DMT for backward DAS and forward IM/DD communication without a sensing-only interblock guard. A CP spanning the sensing-channel memory establishes circular convolution after CP removal, enabling one-tap sensing-channel FDE and exact grid-aligned channel reconstruction under the derived ideal-inversion conditions. The analysis also characterizes the cyclic-redundancy cost and slow-time constraints of the full-memory block structure.
\vspace{2pt}

\item
We further develop NoCP-DMT, which retains cross-boundary range-cell contributions in a long-memory Toeplitz model and jointly estimates the channel taps by regularized LS. The formulation makes explicit that removing the transmitted CP does not remove the RTT-scale physical memory or the identifiability requirements. NoCP-DMT eliminates cyclic redundancy at higher reconstruction complexity, while overlapping observation windows permit a shorter hop and a higher nominal slow-time update rate.
\vspace{2pt}

\item
Experiments over a 10-km link show blind localization of a 600-Hz disturbance applied by a piezoelectric transducer (PZT) near 5.071~km and gauge-differential phase correlations of 0.9982/0.9983 for CP-DMT/NoCP-DMT. Across the 2- and 1-V fifteen-record sets, localization standard deviations are 0.43/0.41~m and 0.38/0.38~m, respectively. The corresponding IM/DD error vector magnitude (EVM) values range from 4.79\% to 5.01\%, with no bit errors. The results jointly validate forward payload recovery and backward DAS for both waveform modes.
\end{itemize}

The rest of this paper is organized as follows. Section~\ref{Princ} establishes the general \(\phi\)-OTDR sensing model and analyzes the guard-interval requirement and spatial ISI of MF-based linear frequency-modulated (LFM) pulse compression. Section~\ref{DMTISAC} develops the proposed continuous DMT framework and presents the CP-DMT and NoCP-DMT channel-reconstruction methods. Section~\ref{ExpDiss} presents the experimental results to validate the feasibility of the proposed ISAC framework. Finally, Section~\ref{Conclu} concludes this paper.

\emph{Notation:}
Bold uppercase symbols denote matrices, e.g., \(\mathbf C_{\tilde{s},s}\) and \(\mathbf T_p\), whereas bold lowercase symbols denote vectors, e.g., \(\mathbf h_p\) and \(\mathbf y_p\). Italic symbols denote scalars, e.g., \(p\), \(N_{\mathrm{cp}}\), and \(T_s\). The sets \(\mathbb{R}\), \(\mathbb{C}\), and \(\mathbb{Z}\) denote the real field, complex field, and set of integers, respectively. The superscripts \((\cdot)^{\mathrm T}\), \((\cdot)^{\mathrm H}\), and \((\cdot)^{*}\) denote transpose, Hermitian transpose, and complex conjugation, respectively. The symbols \(|\cdot|\), \(\|\cdot\|_1\), and \(\|\cdot\|_2\) denote absolute value, the \(\ell_1\)-norm, and the Euclidean norm, respectively.

\section{Phase-Sensitive Optical Time-Domain Reflectometry-Based Distributed Acoustic Sensing}\label{Princ}
In Section~\ref{SysMod}, we develop a general \(\phi\)-OTDR model from coherent optical reception to the discrete finite-memory distributed sensing channel. Section~\ref{ISIByMF} specializes this model to MF-based pulse compression and identifies the guard requirement imposed by RTT and spatial ISI.
\subsection{Phase-Sensitive Optical Time-Domain Reflectometry}\label{SysMod}

\begin{figure}[!t]
    \centering
    \includegraphics[scale=0.74]{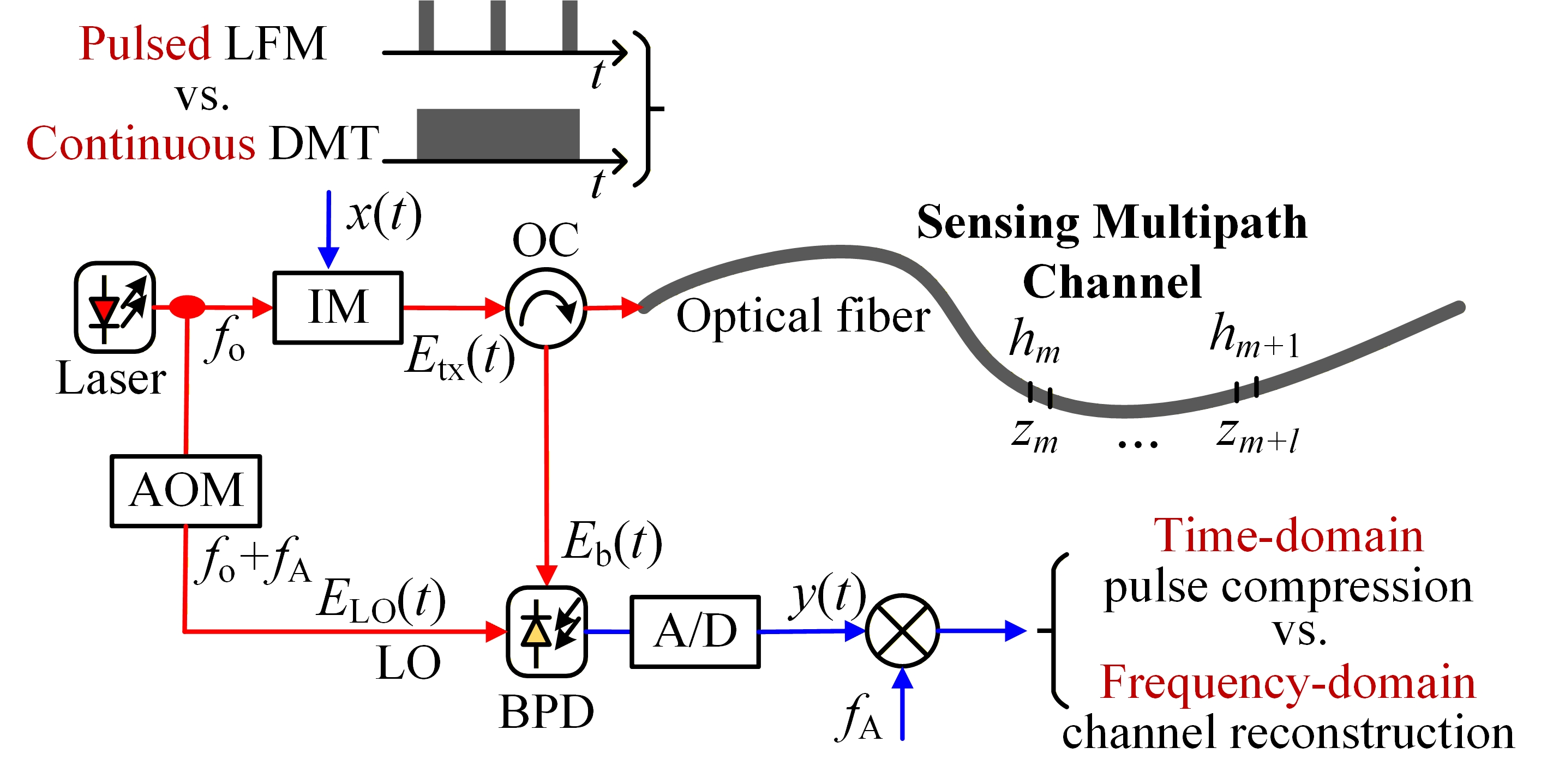}
    \caption{A phase-sensitive optical time-domain reflectometry (\(\phi\)-OTDR) system with intensity modulation and heterodyne coherent reception for distributed acoustic sensing (DAS). The known electrical waveform \(x(t)\) can be a pulsed linear frequency-modulated (LFM) waveform or a continuous payload-bearing discrete multitone (DMT) waveform, with the corresponding return processed through pulse compression or distributed-channel reconstruction, respectively. IM: intensity modulator; OC: optical circulator; AOM: acousto-optic modulator; LO: local oscillator; BPD: balanced photodetector; A/D: analog-to-digital converter.}
    \label{fig1}
\end{figure}

Fig.~\ref{fig1} shows the general architecture of a \(\phi\)-OTDR system for DAS with an intensity modulator (IM) and heterodyne coherent reception.
In the backward DAS branch, an acousto-optic modulator (AOM) frequency shifts the local oscillator (LO); the heterodyne model developed below pertains exclusively to this branch.
The adopted single-branch heterodyne front end enables recovery of the complex backscatter field from the output of a single balanced photodetector (BPD) through digital downconversion. Alternatively, a \(90^\circ\) optical hybrid followed by two BPDs can provide the in-phase and quadrature (I/Q) outputs, yielding the same complex baseband observation after carrier-offset compensation.
In Fig.~\ref{fig1}, the laser at optical frequency \(f_0\) is divided into a probe branch and an LO branch.
In the probe branch, a real-valued electrical waveform \(x(t)\in\mathbb{R}\), constrained by the transmitter bandwidth and drive range, drives the IM.

Let \(e_{\mathrm{tx}}(t)\) denote the complete complex envelope of the launched optical field at the transmitter output, including the effects of the IM, the deterministic transmitter response, and any bias-induced optical-carrier component.
The corresponding launched analytic optical field is written as \(E_{\mathrm{tx}}(t)=e_{\mathrm{tx}}(t)e^{j2\pi f_0t}\).
The launched optical power and its local linear approximation are
\begin{equation}
    P_{\mathrm{tx}}(t)
    =
    |e_{\mathrm{tx}}(t)|^2  \simeq
    \bar{P}_{\mathrm{tx}}\left(1+\mu_{\mathrm{IM}}x(t)\right)  \ge 0,
    \label{eq:im_power}
\end{equation}
where \(\bar{P}_{\mathrm{tx}}\) is the bias-point optical power and \(\mu_{\mathrm{IM}}\) is the effective IM index.
Note that the affine form in \eqref{eq:im_power} is only used over the selected linear operating range.
For sensing reconstruction, let \(e(t)\) denote the known continuous-time effective sensing excitation obtained from the modulation-bearing optical field after deterministic carrier/bias removal and reference calibration.
Any residual bias-carrier contribution not removed by preprocessing is included in the equivalent disturbance \(w(t)\).
The relationship between \(e_{\mathrm{tx}}(t)\) and \(e(t)\) depends on the modulation format and is specified for each probing waveform below.

In the LO branch, the AOM shifts the optical frequency by \(f_{\mathrm A}\).
Let \(A_{\mathrm{LO}}\) denote the complex LO amplitude, and then the LO field is expressed as
\begin{equation}
    E_{\mathrm{LO}}(t)
    =
    A_{\mathrm{LO}}e^{j2\pi(f_0+f_{\mathrm A})t}.
    \label{eq:aom_lo_field}
\end{equation}
Let the Rayleigh-backscattered analytic field at the BPD be \(E_{\mathrm b}(t)=b(t)e^{j2\pi f_0t}\), where \(b(t)\) is its complex envelope.
After removal of the direct-current (DC) component, the intermediate-frequency (IF) LO--backscatter beat current can be represented as
\begin{equation}
    i_{\mathrm{IF}}(t)
    =
    2R_{\mathrm{pd}}\operatorname{Re}
    \left\{
    A_{\mathrm{LO}}^{*}b(t)e^{-j2\pi f_{\mathrm A}t}
    \right\}
    +
    i_{\mathrm{ss}}(t)
    +
    n_i(t),
    \label{eq:heterodyne_current}
\end{equation}
where \(R_{\mathrm{pd}}\) is the photodetector responsivity, \(i_{\mathrm{ss}}(t)\) is the signal--signal beating term, and \(n_i(t)\) is receiver noise.
Balanced detection and LO-dominant heterodyne operation suppress the signal--signal beating term, while beat-band selection and digital filtering reject remaining out-of-band components. Any residual contribution is included in the equivalent disturbance~\cite{LuJLT2010,VidalOE2023Noise}.
With the digital-mixing sign chosen consistently with the retained heterodyne image, complex mixing and low-pass filtering give
\begin{equation}
    y(t)
    =
    \mathcal{L}_{B}\!
    \left\{
    2i_{\mathrm{IF}}(t)e^{j2\pi f_{\mathrm A}t}
    \right\},
    \label{eq:digital_downconversion}
\end{equation}
where \(B\) is the retained complex-baseband reconstruction bandwidth and \(\mathcal{L}_{B}\{\cdot\}\) denotes the corresponding low-pass filtering operation.
After image rejection and absorption of deterministic transceiver gains, we have
\begin{equation}
    y(t)
    =
    \kappa b(t)
    +
    w(t),
    \label{eq:heterodyne_baseband}
\end{equation}
where \(\kappa\) is the effective receiver gain and \(w(t)\) includes receiver noise, residual signal--signal beating, image leakage, and filtering residuals.
Furthermore, the deterministic gain \(\kappa\) is absorbed into the distributed sensing-channel response defined below.
Consequently, \(f_{\mathrm A}\) is introduced solely through the LO branch rather than by frequency translating the launched waveform. Beat-band selection and digital downconversion then yield the complex observation \(y(t)\), whose signal component contains the amplitude and phase information required by the distributed sensing-channel model below.

To model the distributed sensing channel, define \(h(z,t)\) as the time-varying equivalent complex Rayleigh-backscatter channel-response density at position \(z\) and observation time \(t\).
Under the single-scattering approximation, the coherent superposition of distributed backscatter contributions yields~\cite{MasoudiOE2017}
\begin{equation}
    y(t)
    =
    \int_0^L
    h(z,t)e\!\left(t-\tau(z)\right)dz
    +
    w(t),
    \label{eq:tv_cont_backscatter_isac}
\end{equation}
where \(L\) is the fiber length, $\tau(z)=\frac{2z}{v_g}$ is the round-trip delay associated with position \(z\), and $T_{\mathrm{RTT}}=\tau(L)=\frac{2L}{v_g}$ denotes the RTT.
Here, \(v_g=c/n_g\) is the fiber group velocity, \(c\) is the speed of light in vacuum, and \(n_g\) is the group index.
Eq.~\eqref{eq:tv_cont_backscatter_isac} is a continuous-time linear time-varying distributed sensing-channel model.
Through the delay--range mapping \(z=v_g\tau/2\), \eqref{eq:tv_cont_backscatter_isac} has the same input--output form as a causal finite-memory sensing multipath channel in the delay domain, as illustrated in Fig.~\ref{fig1}; the second argument of \(h(z,t)\) describes its temporal evolution under acoustic perturbations.

Assume that the distributed channel-response density remains approximately constant over one observation interval.
Let \(\mathcal{T}_p\) denote the \(p\)-th observation interval and \(t_p\) its representative slow-time instant.
When the duration of \(\mathcal{T}_p\) is short relative to the channel coherence time,
\begin{equation}
    h(z,t)
    \simeq
    h(z,t_p)
    \triangleq
    h_p(z), 
    \label{eq:quasi_static_channel}
\end{equation}
where $t\in\mathcal{T}_p$.
The general time-varying model then reduces within \(\mathcal{T}_p\) to
\begin{equation}
    y_p(t)
    =
    \int_0^L
    h_p(z)e\!\left(t-\tau(z)\right)dz
    +
    w_p(t),
    \label{eq:cont_backscatter_isac}
\end{equation}
where \(y_p(t)\) and \(w_p(t)\) represent the received waveform and equivalent disturbance over the \(p\)-th observation interval.
The input remains the global \(e(t)\), including the excitation segment over the preceding \(T_{\mathrm{RTT}}\) required to represent the channel memory.
Accordingly, \eqref{eq:cont_backscatter_isac} applies to both pulsed and continuous probing.

The channel-response snapshot \(h_p(z)\) incorporates the random Rayleigh-scattering coefficient, two-way propagation attenuation, static round-trip phase, and perturbation-induced phase variation, and is modeled as~\cite{TLSMOTDR}
\begin{equation}
    h_p(z)
    =
    \rho(z)e^{-2\alpha z}e^{-j2\beta_0 z}e^{j\phi_p(z)},
    \label{eq:rayleigh_response}
\end{equation}
where \(\rho(z)\) is the equivalent complex Rayleigh-scattering coefficient, including the range-independent transceiver gain absorbed from \(\kappa\); \(\alpha\) is the fiber amplitude-attenuation coefficient; \(\beta_0\) is the propagation constant at the optical carrier; and \(\phi_p(z)\) is the equivalent perturbation-induced round-trip phase.
Dynamic strain changes \(\phi_p(z)\) over the slow-time index \(p\). Hence, acoustic information is recovered from the evolution of the distributed channel response rather than from a single backscatter record, as discussed in~\cite{JuarezJLT2005,LuJLT2010}.

When the recovered complex-baseband waveform is sampled on the DAS reconstruction grid with interval \(T_s=1/F_s\), the delay-bin spacing is mapped to the range-bin spacing
\begin{equation}
    \Delta z = \frac{v_gT_s}{2}.
    \label{eq:delta_z}
\end{equation}
The \(m\)-th delay grid point corresponds to delay \(mT_s\) and range \(z_m=m\Delta z\).
Let \(t_{p,0}\) denote the time origin of \(\mathcal{T}_p\), and define \(e_p[n]\triangleq e(t_{p,0}+nT_s)\), \(y_p[n]\triangleq y(t_{p,0}+nT_s)\), and \(w_p[n]\triangleq w(t_{p,0}+nT_s)\). The input context \(e_p[n]\) is defined over all indices required by the channel memory and includes samples preceding the retained output window.
Since the fiber occupies only the interval \([0,L]\), define the \(m\)-th range cell as
\begin{equation}
    \mathcal{Z}_m = \left[z_m-\Delta z/2,z_m+\Delta z/2\right) \cap [0,L],
    \label{eq:range_bin_interval}
\end{equation}
and define the corresponding distributed-channel tap as
\begin{equation}
    h_{p,m}
    \triangleq
    \int_{\mathcal{Z}_m}h_p(z)\,dz.
    \label{eq:range_bin_coeff}
\end{equation}
According to~\eqref{eq:cont_backscatter_isac}, approximating the propagation delay within \(\mathcal{Z}_m\) by \(mT_s\) yields
\begin{equation}
    y_p[n]
    =
    \sum_{m=0}^{M-1}
    h_{p,m}e_p[n-m]
    +
    w_p[n],
    \label{eq:discrete_backscatter_isac}
\end{equation}
where \(M_{\min}\triangleq\lceil T_{\mathrm{RTT}}/T_s\rceil+1\) is the minimum RTT-covering support order. The reconstruction model may use \(M\ge M_{\min}\); taps beyond the physical fiber support are zero in the ideal fiber-only model.
Eq.~\eqref{eq:discrete_backscatter_isac} is the unified discrete finite-memory sensing-channel model, with \(\mathbf h_p\triangleq[h_{p,0},\ldots,h_{p,M-1}]^{\mathrm T}\). Each nonzero tap coherently aggregates the Rayleigh backscatter from a physical range cell. The indices \(n\), \(m\), and \(p\) denote received sample, propagation-delay/range, and slow-time snapshot, respectively; range is recovered from \(m\). Waveform formats change the known input and reconstruction operation, but not this linear-convolution model.

\subsection{Phase-Sensitive Optical Time-Domain Reflectometry Using MF-Based LFM Pulse Compression}\label{ISIByMF}
MF-based pulse compression in \(\phi\)-OTDR improves the energy--spatial-resolution tradeoff by transmitting a long-duration broadband pulse and applying matched filtering to the received trace~\cite{ZouOE2015,MompoOL2018}.
Let \(s_{\mathrm{LFM}}(t)\) denote the complex representation of an LFM pulse of duration \(T_{\mathrm{LFM}}\), whose real electrical modulation waveform is \(x_{\mathrm{LFM}}(t)=\operatorname{Re}\!\left\{s_{\mathrm{LFM}}(t)\right\}\).
For pulsed LFM operation, the IM is additionally gated toward its optical off state outside each LFM pulse. Thus, the launched optical field, rather than only the nominal electrical LFM component, is negligible during the guard interval.
Let \(s[n]\) denote the sampled complex-baseband MF reference, and let \(\tilde{s}[n]\) denote the corresponding effective LFM excitation on the DAS reconstruction grid, including the deterministic transmitter mapping used to define \(e_p[n]\).
Under the finite-support pulse model, residual out-of-window probing energy is assumed negligible, and \(\tilde{s}[n]\) is idealized as zero for \(n\notin\{0,\ldots,N_s-1\}\).
Thus, \(N_s\) denotes the support length of the effective LFM excitation rather than that of the nominal electrical sequence.

With a pulse-repetition interval (PRI) of \(N_{\mathrm{PRI}}\) samples and integer pulse index \(q\in\mathbb{Z}\), the discrete LFM pulse train is
\begin{equation}
    e_{\mathrm{LFM}}[n]
    =
    \sum_{q\in\mathbb{Z}}
    \tilde{s}[n-qN_{\mathrm{PRI}}].
    \label{eq:lfm_pulse_train}
\end{equation}
For this pulse-specific model, let \(\mathcal{T}_p\) denote the observation interval aligned with the start of the \(p\)-th PRI, with \(t_{p,0}=pN_{\mathrm{PRI}}T_s\), and let \(y_p[n]\) denote the corresponding complete return record. The quasi-static approximation is imposed over \(\mathcal{T}_p\).
The required input context is then
\begin{align}
    e_{\mathrm{LFM},p}[n]
    &=
    e_{\mathrm{LFM}}[pN_{\mathrm{PRI}}+n] \notag\\
    & =
    \sum_{q\in\mathbb{Z}}
    \tilde{s}[n+(p-q)N_{\mathrm{PRI}}].
    \label{eq:lfm_pulse_train_rx}
\end{align}
Thus, the input \(e_p[n]\) in \eqref{eq:discrete_backscatter_isac} specializes to \(e_{\mathrm{LFM},p}[n]\) for pulsed LFM operation.
Substituting \eqref{eq:lfm_pulse_train_rx} into \eqref{eq:discrete_backscatter_isac}, separating the desired \(q=p\) contribution, and identifying the remaining \(q\ne p\) contributions as IPI gives
\begin{align}
    y_p[n]
    ={}&
    \sum_{m=0}^{M-1}
    h_{p,m}\tilde{s}[n-m]
    \notag\\
    &+
    \underbrace{
    \sum_{\substack{q\in\mathbb{Z}\\q\ne p}}
    \sum_{m=0}^{M-1}
    h_{p,m}\tilde{s}[n+(p-q)N_{\mathrm{PRI}}-m]
    }_{\text{Inter-pulse interference}}
    +w_p[n].
    \label{eq:pulse_ipi_model}
\end{align}
The convolution of an \(N_s\)-sample pulse with an \(M\)-tap distributed sensing channel has support contained within \(N_s+M-1\) samples.
Accordingly, the complete return associated with pulse \(p\) is contained in \(0\le n\le N_s+M-2\), and adjacent returns have disjoint supports if
\begin{equation}
    N_{\mathrm{PRI}}
    \ge
    N_s+M-1,
    \label{eq:pulse_guard_condition}
\end{equation}
where \(N_{\mathrm g}\triangleq N_{\mathrm{PRI}}-N_s\ge M-1\) is the number of silent guard samples between the end of one effective pulse and the start of the next.
Using the sample-interval convention, \(T_{\mathrm{LFM}}=N_sT_s\), \(T_{\mathrm{PRI}}=N_{\mathrm{PRI}}T_s\), and \(T_{\mathrm g}=N_{\mathrm g}T_s\).
As a result, the corresponding continuous-time sufficient condition is \(T_{\mathrm{PRI}}\ge T_{\mathrm{LFM}}+T_{\mathrm{RTT}}\), or, equivalently, \(T_{\mathrm g}\ge T_{\mathrm{RTT}}\).

Under the finite-support model and assuming negligible residual effective LFM excitation outside the pulse, the IPI term in \eqref{eq:pulse_ipi_model} is zero over the complete return record when \eqref{eq:pulse_guard_condition} holds.
The required guard interval therefore scales with fiber length and constrains the pulse-repetition rate~\cite{HartogBook2017}.
Under this isolated-pulse condition,  \eqref{eq:pulse_ipi_model} reduces to
\begin{equation}
    y_p[n]
    =
    \sum_{m=0}^{M-1}
    h_{p,m}\tilde{s}[n-m]
    +
    w_p[n].
    \label{eq:pulse_rx_isac}
\end{equation}
Following standard MF processing of an LFM waveform~\cite{LevanonRadarSignals,RichardsRadar}, the output at range bin \(\ell\) is
\begin{equation}
    r_p[\ell]
    =
    \sum\nolimits_n
    y_p[n]s^{*}[n-\ell],
    \label{eq:mf_def_isac}
\end{equation}
Substituting \eqref{eq:pulse_rx_isac} into \eqref{eq:mf_def_isac} gives
\begin{align}
    r_p[\ell]
    &=
    \sum_{m=0}^{M-1}
    h_{p,m}\!
    \left(\!\sum_n \tilde{s}[n-m]s^{*}[n-\ell]\!\right)
    +
    v_p[\ell] \notag\\
    &=
    \sum_{m=0}^{M-1}
    h_{p,m}R_{\tilde{s},s}[\ell-m]
    +
    v_p[\ell],
    \label{eq:mf_conv_isac}
\end{align}
where \(R_{\tilde{s},s}[q]\triangleq\sum_n \tilde{s}[n]s^{*}[n-q]\) is the aperiodic cross-correlation between the effective LFM excitation and the MF reference, and \(v_p[\ell]\triangleq\sum_n w_p[n]s^{*}[n-\ell]\) is the filtered disturbance.
If \(\tilde{s}[n]=s[n]\), then \(R_{\tilde{s},s}[q]\) reduces to the LFM autocorrelation.
Therefore, the MF output alone does not invert the distributed sensing channel. Instead, the channel is convolved with a waveform-dependent correlation kernel.

Assume that deterministic delay alignment places the correlation peak at zero lag, and define \(\alpha_s\triangleq R_{\tilde{s},s}[0]\ne0\) and \(c_{\tilde{s},s}[q]\triangleq R_{\tilde{s},s}[q]/\alpha_s\).
The normalized MF-based range-bin channel estimate is
\begin{align}
    \widehat{h}^{\mathrm{MF}}_{p,\ell}
    &=
    \frac{r_p[\ell]}{\alpha_s}
    \notag\\
    &=
    h_{p,\ell}
    +
    \underbrace{
    \sum_{\substack{m=0\\m\ne \ell}}^{M-1}
    h_{p,m}c_{\tilde{s},s}[\ell-m]
    }_{\text{Spatial intersymbol interference}}
    +
    \widetilde{v}_p[\ell],
    \label{eq:mf_est_isac}
\end{align}
where \(\widetilde{v}_p[\ell]\triangleq v_p[\ell]/\alpha_s\).
Since \(c_{\tilde{s},s}[0]=1\), the \(m=\ell\) term equals the desired channel tap \(h_{p,\ell}\).
The remaining terms couple the distributed-channel taps associated with other range bins into bin \(\ell\), referred to here as spatial ISI.
Equivalently, let \(\widehat{\mathbf h}^{\mathrm{MF}}_p\) collect the MF-based range-bin channel estimates. Then
\begin{equation}
    \widehat{\mathbf h}^{\mathrm{MF}}_p
    =
    \mathbf C_{\tilde{s},s}\mathbf h_p
    +
    \widetilde{\mathbf v}_p,
    \label{eq:mf_matrix_isac}
\end{equation}
where \(\widetilde{\mathbf v}_p=[\widetilde v_p[0],\ldots,\widetilde v_p[M-1]]^{\mathrm T}\), and \(\mathbf C_{\tilde{s},s}\in\mathbb{C}^{M\times M}\) has entries \([\mathbf C_{\tilde{s},s}]_{\ell,m}=c_{\tilde{s},s}[\ell-m]\) for \(0\le\ell,m\le M-1\).
The diagonal entries of \(\mathbf C_{\tilde{s},s}\) are unity, whereas its off-diagonal entries are the spatial ISI coupling coefficients.
The inter-range-bin coupling operator is identically absent for arbitrary channel vectors only if \(c_{\tilde{s},s}[q]=\delta[q]\) over the represented tap differences, where \(\delta[q]\) is the Kronecker delta.

Practical finite-duration, finite-bandwidth LFM pulses generally have nonzero off-peak correlation samples~\cite{LevanonRadarSignals,RichardsRadar}.
Windowing can reduce these sidelobes but broadens the compressed main lobe and therefore degrades spatial resolution~\cite{MompoOL2018}.
Accordingly, leakage from a strong range-bin response can mask a weaker response in a neighboring bin, produce spurious spatial components, elevate the background, or distort the recovered phase trace.
Increasing the sampling rate at a fixed occupied bandwidth only provides a denser sampling of the same bandwidth-limited correlation response and does not eliminate the underlying spatial ISI coupling.
\vspace{2pt}

After channel reconstruction, a \(\phi\)-OTDR receiver recovers the gauge-differential phase from the complex range response \(\widehat h^{\mathrm d}_{p,\ell}\)~\cite{GabaiOL2016}. For MF reception, \(\widehat h^{\mathrm d}_{p,\ell}=\widehat h^{\mathrm{MF}}_{p,\ell}\). For the continuous DMT modes developed in Section~\ref{DMTISAC}, frequency-diversity combining precedes phase recovery to mitigate Rayleigh fading~\cite{ChenPCC2026}. The replicas \(\widehat h^{(v)}_{p,\ell}=\mathcal F^{-1}\{W_v[k]\mathcal H_p[k]\}[\ell]\) are obtained from the reconstructed channel spectrum \(\mathcal H_p[k]\), where \(\mathcal F^{-1}\) denotes the inverse discrete Fourier transform. Here, \(W_0[k]\) retains the full reconstruction band, whereas \(W_v[k]\), \(v=1,\ldots,V\), are overlapping raised-cosine windows. The replicas share the range grid and have correlated noise.

Stack the replicas into \(\mathbf r_{p,\ell}\). For noise-whitened principal component combining (PCC), let \(\mathbf L\mathbf L^{\mathrm H}=\mathbf R_n+\epsilon_n\mathbf I\), where \(\mathbf R_n\) is the replica-noise covariance matrix with a unit diagonal average and \(\epsilon_n>0\). The dominant unit-norm eigenvector \(\mathbf u_\ell\) of the uncentered covariance of \(\mathbf L^{-1}\mathbf r_{p,\ell}\), estimated over an initial reliable calibration interval, gives
\begin{equation}
    \widehat h^{\mathrm d}_{p,\ell}
    =\frac{\mathbf u_\ell^{\mathrm H}\mathbf L^{-1}\mathbf r_{p,\ell}}
    {\|\mathbf L^{-\mathrm H}\mathbf u_\ell\|_1}.
    \label{eq:pcc_channel}
\end{equation}
The combining weights remain fixed over the record. Under an approximately common perturbation phase across the spatial responses of the replicas, the combining weights in~\eqref{eq:pcc_channel} preserve the temporal evolution up to a static offset.

For either receiver, a gauge separation of \(G\) samples gives
\begin{equation}
    \widehat g_{p,\ell}
    =\widehat h^{\mathrm d}_{p,\ell+G}
    \bigl(\widehat h^{\mathrm d}_{p,\ell}\bigr)^*,
    \label{eq:gauge_product}
\end{equation}
with gauge length \(G\Delta z\) and center coordinate \(z_\ell+G\Delta z/2\). After relative endpoint-amplitude and phase-continuity screening, each reliable continuous interval \(\mathcal I\), starting at \(p_0\), yields
\begin{equation}
    \widehat\Phi_{p,\ell}
    =\operatorname{unwrap}_{p\in\mathcal I}
    \!\left\{\angle\widehat g_{p,\ell}\right\}
    -\angle\widehat g_{p_0,\ell}.
    \label{eq:gauge_phase}
\end{equation}
Here, \(\angle\{\cdot\}\in(-\pi,\pi]\) denotes the principal phase, and unwrapping acts along slow time without imposing continuity across unreliable intervals. For MF reception, the spatial ISI in \eqref{eq:mf_est_isac} can therefore produce phase bias and slow-time waveform distortion.

MF-based LFM pulse-compression \(\phi\)-OTDR therefore involves two distinct limitations.
An RTT-scale silent guard interval is required to prevent overlap between the returns of different transmitted pulses.
With one distributed-channel snapshot reconstructed per PRI, \(1/T_{\mathrm{PRI}}\) is also the slow-time sampling rate; hence, under uniform sampling, the maximum unaliased acoustic frequency is bounded by \(1/(2T_{\mathrm{PRI}})\)~\cite{HartogBook2017,MompoOL2018}.
Even when this isolated-pulse condition is satisfied, the off-peak correlation samples couple different distributed-channel taps within the return from the same pulse, resulting in spatial ISI.
The IPI-avoidance condition is therefore controlled by \(T_{\mathrm{PRI}}\), which introduces a tradeoff between return isolation and alias-free slow-time bandwidth, whereas spatial ISI is governed by the pulse-compression kernel and cannot be removed by extending the guard interval.

In a waveform-sharing ISAC link, however, preserving return isolation requires this RTT-scale silent guard interval to remain free of the shared transmitted waveform; it therefore carries no payload and reduces the payload duty cycle.
The loss can become substantial when \(T_{\mathrm{RTT}}\) dominates the useful waveform duration, making periodically gated probing poorly suited to continuous, high-duty-cycle data delivery.
The next section therefore replaces the pulsed LFM waveform with a continuous payload-bearing DMT waveform and replaces MF correlation with explicit distributed-channel reconstruction before phase extraction. For backward DAS, the same transmitted waveform provides the effective DMT excitation used for distributed-channel reconstruction, while also supporting forward IM/DD payload recovery.

\section{Continuous Discrete Multitone Modulation Framework for ISAC}\label{DMTISAC}
In Section~\ref{DMTFDE}, we develop the CP-DMT sensing model, establish the conditions under which a full-memory CP yields circular convolution and spatial-ISI-free one-tap FDE, and characterize the associated overhead and slow-time constraints. Section~\ref{NoCPDMT} develops a finite-memory Toeplitz sensing model for NoCP-DMT and establishes regularized LS reconstruction, identifiability, and overlapping-window slow-time conditions. In either mode, one continuous payload-bearing DMT waveform supports forward communication and backward DAS.

\subsection{Phase-Sensitive Optical Time-Domain Reflectometry Using Frequency-Domain Channel Reconstruction}\label{DMTFDE}
\begin{figure}[!t]
    \centering
    \includegraphics[scale=0.9]{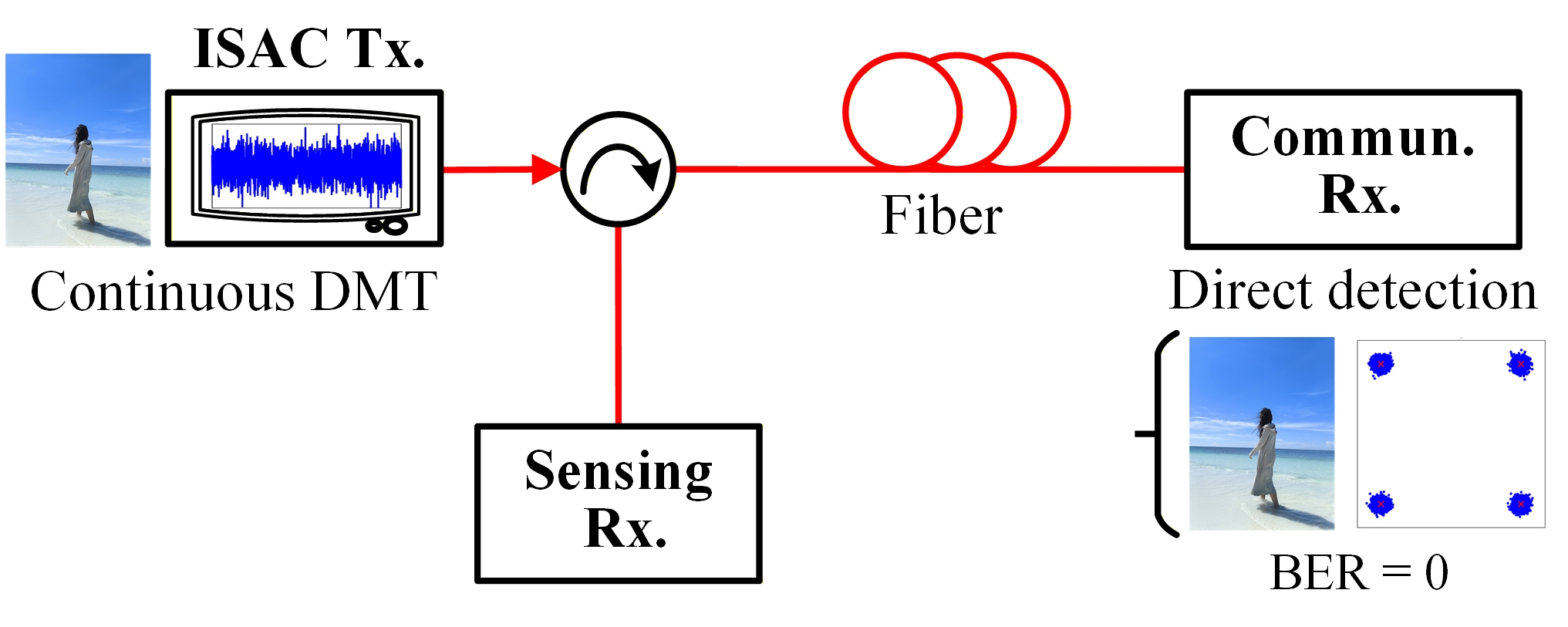}
    \caption{Illustration of a single continuous discrete multitone (DMT) waveform for fiber-optic integrated sensing and communication (ISAC), where the forward payload is recovered by an intensity-modulation/direct-detection (IM/DD) DMT communication receiver, and the backward Rayleigh-scattered field is processed by the coherent sensing receiver.}
    \label{fig2}
\end{figure}

The CP-DMT branch uses frequency-domain channel reconstruction for backward DAS, while the forward branch in Fig.~\ref{fig2} is a directly detected IM/DD DMT link with pilot-aided one-tap equalization after CP removal.

Heterodyne detection recovers the Rayleigh backscatter using the known transmit-derived reference for each CP-DMT block under the adopted linearized IM model. Successive payload-bearing blocks are transmitted continuously without a sensing-only interblock guard, replacing the pulse-and-wait operation of Section~\ref{ISIByMF}.

In this section, \(p\) indexes both a transmitted CP-DMT block and the distributed-channel snapshot associated with that block. Let \(S_p[k]\), \(k=0,\ldots,N-1\), denote the subcarrier coefficients of the \(p\)-th payload-bearing useful DMT symbol. For even \(N\), imposing \(S_p[N-k]=S_p^{*}[k]\), \(k=1,\ldots,N/2-1\), with real-valued DC and Nyquist coefficients gives the real-valued \(N\)-sample useful DMT symbol~\cite{ArmstrongJLT2009,ShiehOFDMBook}
\begin{equation}
    u_p[n]
    =
    \frac{1}{N}
    \sum_{k=0}^{N-1}
    S_p[k]e^{j2\pi kn/N},
    \label{eq:dmt_useful}
\end{equation}
where $0\le n\le N-1$.
The corresponding CP-DMT block is
\begin{equation}
    u_{p,\mathrm{cp}}[n]
    =
    u_p[(n)_N],
    \label{eq:dmt_cp}
\end{equation}
where \((n)_N\) denotes the modulo-\(N\) index, \(N_{\mathrm{cp}}\) is the CP length, and $-N_{\mathrm{cp}}\le n\le N-1$.

To represent the effective sensing excitation in the discrete channel model, let \(q_{p,\mathrm{cp}}[n]\) denote the known transmit-derived reconstruction reference for the \(p\)-th CP-DMT block on the DAS reconstruction grid under the linearized quadrature-biased IM model.
This reconstruction reference is constructed from the final normalized digital waveform and is a proportional proxy for the modulation-bearing component of the launched optical field, rather than a measured post-IM field.
Deterministic linear shaping accounted for by calibration of the reconstruction reference is included in \(q_{p,\mathrm{cp}}[n]\), whereas the remaining unknown range-independent gain and sign are absorbed into the reconstructed distributed sensing channel.
The circular-convolution model imposes the cyclic extension
\begin{equation}
    q_{p,\mathrm{cp}}[n]
    =
    q_p[(n)_N],
    \label{eq:dmt_q_cp_condition}
\end{equation}
where $-N_{\mathrm{cp}}\le n\le N-1$ and \(q_p[n]\triangleq q_{p,\mathrm{cp}}[n]\) for \(0\le n\le N-1\). Thus, \(q_{p,\mathrm{cp}}[n]\) serves as the known reconstruction reference for the effective sensing excitation in the distributed sensing-channel model, with \eqref{eq:dmt_q_cp_condition} imposed on the aligned reconstruction grid.

Let \(N_{\mathrm b}\triangleq N+N_{\mathrm{cp}}\). The global sampled excitation sequence of the continuous CP-DMT stream satisfies
\begin{equation}
    e_{\mathrm{DMT}}[pN_{\mathrm b}+N_{\mathrm{cp}}+n]
    =
    q_{p,\mathrm{cp}}[n],
    \label{eq:continuous_cp_dmt_stream}
\end{equation}
where $-N_{\mathrm{cp}}\le n\le N-1$.
Successive CP-DMT blocks indexed by \(p\) can carry different payload data. Every transmitted sample belongs to either CP or useful symbols of a CP-DMT block. No zero-valued interblock guard interval is inserted.

Let \(y_{\mathrm{DMT}}[n]\) denote the global received sequence corresponding to \(e_{\mathrm{DMT}}[n]\). For the \(p\)-th distributed-channel snapshot, choose the local time origin at the beginning of its useful DMT symbol and define the retained samples as
\begin{equation}
    z_p[n]
    \triangleq
    y_{\mathrm{DMT}}[pN_{\mathrm b}+N_{\mathrm{cp}}+n],
    \label{eq:dmt_retained_samples}
\end{equation}
where $0\le n\le N-1$.
Substituting \eqref{eq:continuous_cp_dmt_stream} into \eqref{eq:discrete_backscatter_isac} gives, before imposing a CP-length condition,
\begin{equation}
    z_p[n]
    = 
    \sum_{m=0}^{\min\{M-1,n+N_{\mathrm{cp}}\}}
    h_{p,m}q_p[(n-m)_N]
    +\iota_p[n]+w_p[n],
    \label{eq:dmt_linear_rx}
\end{equation}
where the interblock interference is expressed as
\begin{equation}
    \iota_p[n]
    \triangleq
    \sum_{m=n+N_{\mathrm{cp}}+1}^{M-1}
    h_{p,m}
    e_{\mathrm{DMT}}[pN_{\mathrm b}+N_{\mathrm{cp}}+n-m].
    \label{eq:dmt_ibi}
\end{equation}
The sum in \eqref{eq:dmt_ibi} is empty when its lower limit exceeds \(M-1\). The terms in the sum require samples from before the CP of the current CP-DMT block and therefore originate from an earlier CP-DMT block.

If
\begin{equation}
    N_{\mathrm{cp}}\ge M-1,
    \label{eq:cp_condition}
\end{equation}
\(\iota_p[n]=0\) for every retained sample. The CP supplies the current-block samples required by the channel memory, while the preceding-block contribution is confined to the discarded CP interval. Provided also that \(M\le N\), CP removal makes the linear convolution over the retained useful DMT symbol equivalent to the \(N\)-point circular convolution
\begin{equation}
    z_p[n]
    =
    \sum_{m=0}^{M-1}
    h_{p,m}q_p[(n-m)_N]
    +w_p[n].
    \label{eq:dmt_circular_rx}
\end{equation}
Thus, the CP does not make the physical propagation globally circular; it establishes a circular model only over each CP-removed useful DMT symbol. With deterministic transceiver shaping accounted for in the reconstruction reference, \(M\) is dominated by the fiber round-trip memory, and \eqref{eq:cp_condition} is a full-memory CP requirement. If residual transceiver memory is retained in the effective distributed sensing channel, both \(M\) and the required CP must instead cover the resulting composite memory.

Define the \(N\)-point discrete Fourier transforms (DFTs)
\begin{equation}
    Q_p[k]
    =
    \sum_{n=0}^{N-1}
    q_p[n]e^{-j2\pi kn/N},
    \label{eq:dmt_fde_reference}
\end{equation}
and
\begin{equation}
    H_p[k]
    =
    \sum_{m=0}^{M-1}
    h_{p,m}e^{-j2\pi km/N}.
    \label{eq:dmt_channel_response}
\end{equation}
The corresponding output and disturbance spectra are 
\begin{align}
    Z_p[k]
    &\triangleq
    \sum_{n=0}^{N-1}z_p[n]e^{-j2\pi kn/N},
    \label{eq:dmt_output_spectra11} \\
    W_p[k]
    &\triangleq
    \sum_{n=0}^{N-1}w_p[n]e^{-j2\pi kn/N}.
    \label{eq:dmt_output_spectra}
\end{align}
Applying the DFT to \eqref{eq:dmt_circular_rx} diagonalizes the circulant channel operator and gives
\begin{equation}
    Z_p[k]
    =
    Q_p[k]H_p[k]+W_p[k].
    \label{eq:dmt_freq_model}
\end{equation}
Let \(\chi_p[k]\in\{0,1\}\) denote the reconstruction mask, and define the retained-subcarrier set as \(\mathcal{K}_p\triangleq\{k\in\{0,\ldots,N-1\}:\chi_p[k]=1\}\). The retained set is selected such that \(Q_p[k]\ne0\) for every \(k\in\mathcal{K}_p\). Regularized one-tap sensing-channel FDE is then defined over all DFT bins as
\begin{equation}
    \widehat H_p^{\mathrm{FDE}}[k]
    =
    \begin{cases}
    \displaystyle
    \frac{Q_p^{*}[k]}
         {|Q_p[k]|^2+\lambda}
    Z_p[k], & k\in\mathcal{K}_p,\\[2mm]
    0, & k\notin\mathcal{K}_p,
    \end{cases}
    \label{eq:dmt_fde}
\end{equation}
where \(\lambda\ge0\). The distributed sensing channel is then reconstructed by the inverse DFT (IDFT) as
\begin{equation}
    \widehat h^{\mathrm{FDE}}_{p,\ell}
    =
    \frac{1}{N}
    \sum_{k=0}^{N-1}
    \widehat H_p^{\mathrm{FDE}}[k]e^{j2\pi k\ell/N},
    \label{eq:dmt_ifft}
\end{equation}
where $0\le\ell\le N-1$.

\begin{figure}[!t]
    \centering
    \includegraphics[width=0.45\textwidth]{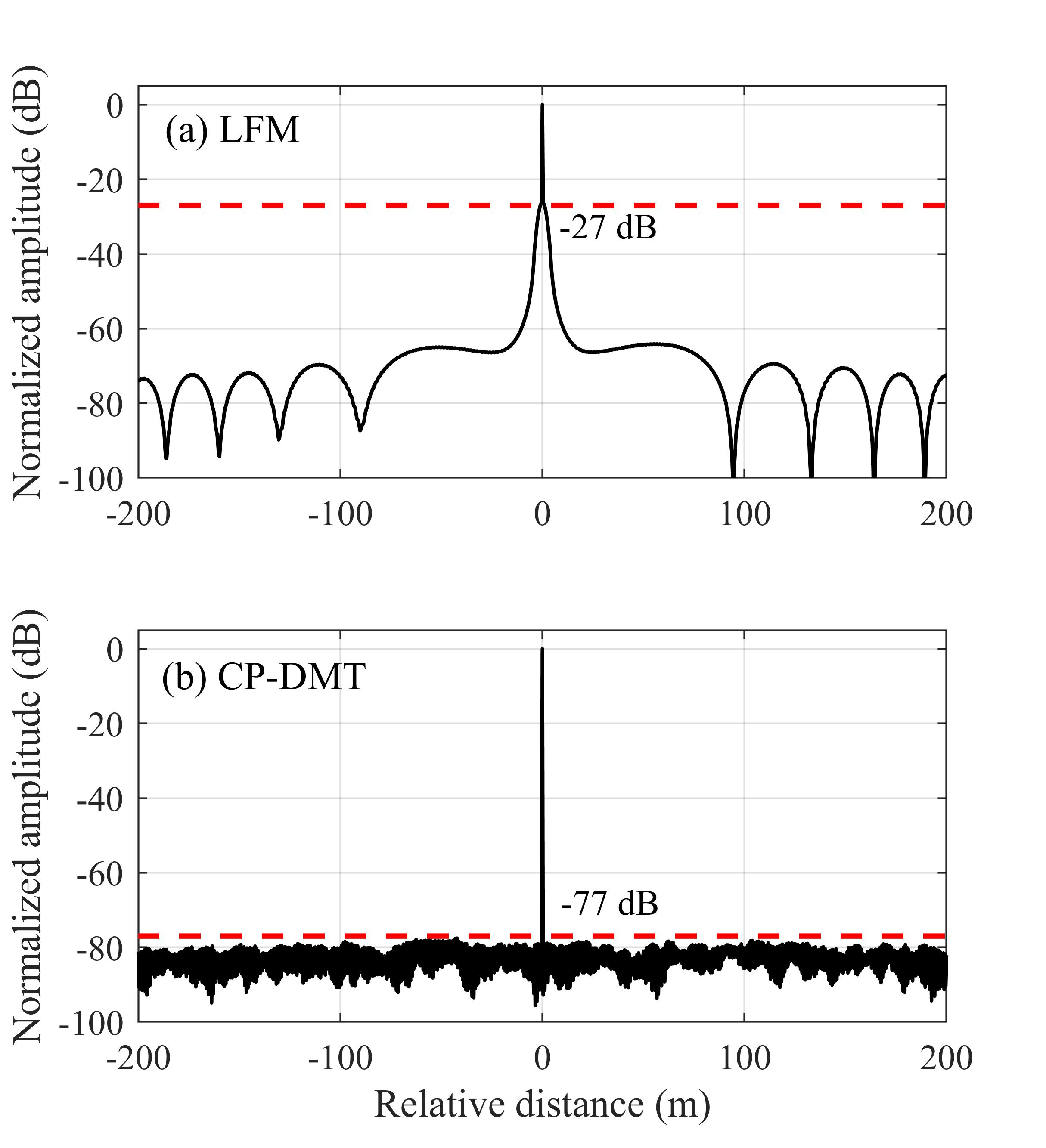}
    \caption{Normalized noise-free end-to-end numerical point-spread functions for a grid-aligned point scatterer using (a) matched-filter-based linear frequency modulation (LFM) pulse compression and (b) frequency-domain cyclic-prefix discrete multitone (CP-DMT) channel reconstruction. The dashed lines indicate rounded off-mainlobe reference levels; panel (b) includes spectral selection, missing-bin interpolation, and small regularization.}
    \label{Resfig1}
\end{figure}

Under sufficient CP, \(M\le N\), exact alignment to the CP-DMT block boundaries, a distributed sensing channel that is quasi-static over each CP-DMT block, and the noiseless, unregularized full-bin conditions \(\mathcal{K}_p=\{0,\ldots,N-1\}\), \(Q_p[k]\ne0\) for all \(k\), and \(\lambda=0\), \eqref{eq:dmt_fde} exactly cancels the known reconstruction-reference spectrum. Substitution into \eqref{eq:dmt_ifft} gives
\begin{align}
    \widehat h^{\mathrm{FDE}}_{p,\ell}
    &=
    \sum_{m=0}^{M-1}
    h_{p,m}
    \left[
    \frac{1}{N}
    \sum_{k=0}^{N-1}
    e^{j2\pi k(\ell-m)/N}
    \right]
    \label{eq:dmt_isi_free_result1}\\
    &=
    \sum_{m=0}^{M-1}
    h_{p,m}\delta_N[\ell-m]
    \label{eq:dmt_isi_free_result11}\\
    &=
    \begin{cases}
    h_{p,\ell}, & 0\le\ell\le M-1,\\
    0, & M\le\ell\le N-1,
    \end{cases}
    \label{eq:dmt_isi_free_result}
\end{align}
where \(\delta_N[\cdot]\) is the periodic Kronecker delta. Hence, the stated sufficient-CP, noiseless, unregularized full-bin conditions yield exact grid-aligned tap recovery without MF-correlation-induced spatial ISI. This analytical result is distinct from the noise-free end-to-end numerical response in Fig.~\ref{Resfig1}(b), which retains practical spectral selection and regularization.
By contrast, \eqref{eq:mf_est_isac} retains the off-peak correlation terms \(c_{\tilde{s},s}[\ell-m]\), as illustrated in Fig.~\ref{Resfig1}~(a).

The ideal relation in \eqref{eq:dmt_isi_free_result} must be distinguished from a practical bandwidth-limited and regularized reconstruction. Using the reconstruction mask defined above, let
\begin{equation}
    A_p[k]
    =
    \begin{cases}
    \displaystyle
    \frac{|Q_p[k]|^2}
         {|Q_p[k]|^2+\lambda}, & k\in\mathcal{K}_p,\\[2mm]
    0, & k\notin\mathcal{K}_p.
    \end{cases}
    \label{eq:dmt_reconstruction_gain}
\end{equation}
The corresponding range-domain estimate can be written as
\begin{equation}
    \widehat h^{\mathrm{FDE}}_{p,\ell}
    =
    \sum_{m=0}^{M-1}
    h_{p,m}a_p[(\ell-m)_N]
    +\widetilde{w}_p[\ell],
    \label{eq:dmt_practical_kernel}
\end{equation}
where
\begin{equation}
    a_p[r]
    =
    \frac{1}{N}
    \sum_{k=0}^{N-1}
    A_p[k]e^{j2\pi kr/N},
    \label{eq:dmt_kernel}
\end{equation}
is the periodic reconstruction kernel, and \(\widetilde{w}_p[\ell]\) is the range-domain disturbance. With all bins retained, \(Q_p[k]\ne0\), and \(\lambda=0\), \(a_p[r]=\delta_N[r]\), reducing \eqref{eq:dmt_practical_kernel} to \eqref{eq:dmt_isi_free_result}. Inactive or unreliable subcarriers, finite reconstruction bandwidth, and regularization instead broaden \(a_p[r]\), as illustrated numerically in Fig.~\ref{Resfig1}(b). Let \(\mathcal{F}_p=\{f_k:k\in\mathcal{K}_p\}\) denote the retained unwrapped complex-baseband frequencies and define \(B_{\mathrm{rec}}\triangleq\max\mathcal{F}_p-\min\mathcal{F}_p\), the total lowest-to-highest retained span rather than a one-sided bandwidth. For an approximately contiguous band, \(\Delta z_{\mathrm{res}}\approx v_g/(2B_{\mathrm{rec}})\)~\cite{HuangArxiv2026}.
The spectrum \(\mathcal H_p[k]=\widehat H_p^{\mathrm{FDE}}[k]\) supplies the replicas for PCC and gauge-differential phase recovery in \eqref{eq:pcc_channel}--\eqref{eq:gauge_phase}.

Continuous CP-DMT transmission replaces the RTT-scale silent guard interval of pulsed LFM operation with cyclic redundancy. It supports uninterrupted transmission and confines interblock interference to discarded samples under \eqref{eq:cp_condition}, but does not eliminate slow-time sampling constraints or the full-memory CP overhead. When one distributed-channel snapshot is reconstructed from each CP-DMT block, the block duration is also the slow-time sampling interval:
\begin{equation}
    T_{\mathrm{blk}}
    =
    (N+N_{\mathrm{cp}})T_s,
    \label{eq:dmt_slow_time}
\end{equation}
where \(f_{\mathrm{slow}}=1/T_{\mathrm{blk}}\). Let \(f_{\mathrm{ac}}\) denote the frequency of an acoustic component. Under uniform slow-time sampling, a necessary no-aliasing condition is \(f_{\mathrm{ac}}<f_{\mathrm{slow}}/2=1/(2T_{\mathrm{blk}})\). This Nyquist condition alone does not determine the usable acoustic bandwidth: the finite block aperture shapes the temporal response, and the distributed channel must remain approximately quasi-static over the reconstruction block. The usable acoustic band is therefore jointly limited by slow-time aliasing, the block-aperture response, and the quasi-static-channel condition. The CP sample-utilization factor is
\begin{equation}
    \eta_{\mathrm{cp}} = \frac{N}{N+N_{\mathrm{cp}}},
    \label{eq:dmt_cp_efficiency}
\end{equation}
which quantifies the fraction of transmitted time-domain samples assigned to the nonredundant useful DMT symbol. Because \(N_{\mathrm{cp}}T_s\) must cover the fiber round-trip memory, this overhead can remain substantial for long-range DAS. Continuous CP-DMT therefore provides a low-complexity sensing-channel FDE route for waveform-level ISAC, while the full-memory CP overhead motivates channel reconstruction without a repeated full-memory CP, which is developed in the next section.

\subsection{Cyclic-Prefix-Free Continuous Discrete Multitone Modulation Waveform}\label{NoCPDMT}
\begin{figure}[!t]
    \centering
    \includegraphics[scale=0.96]{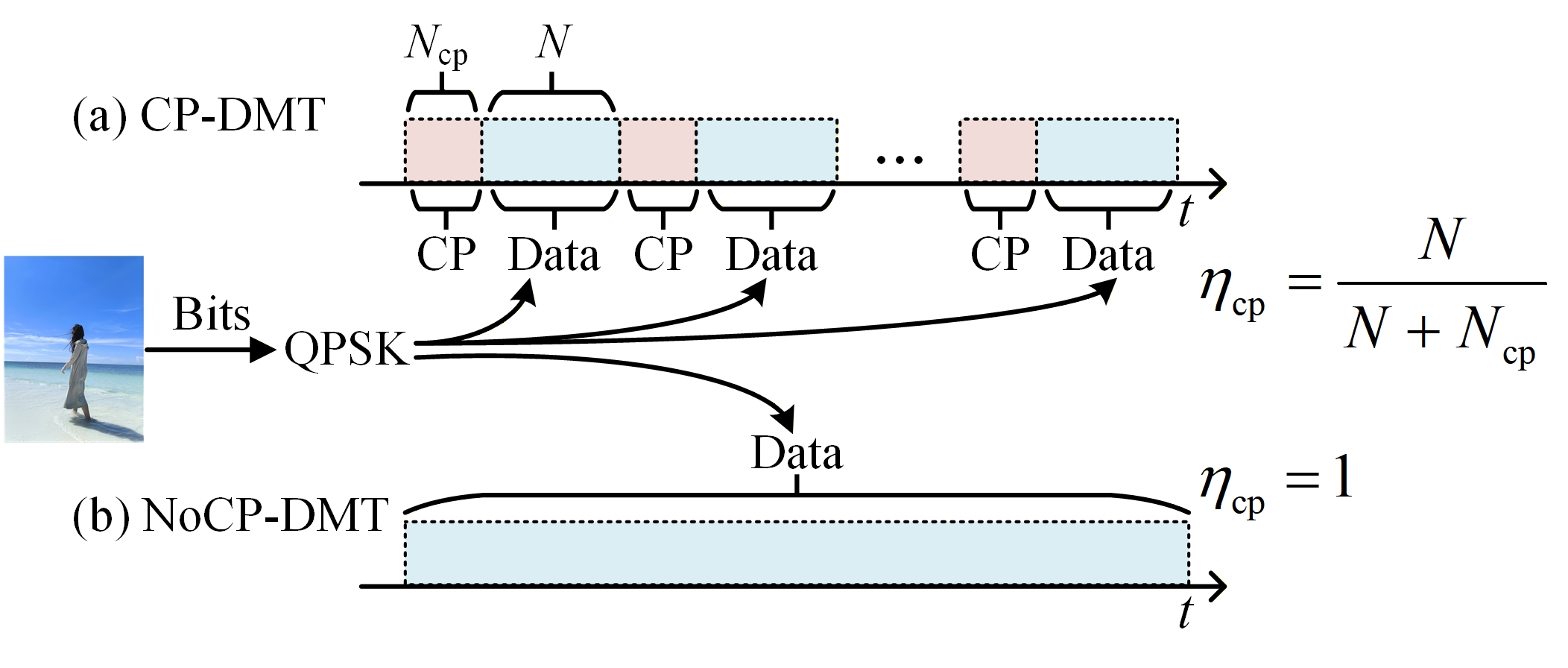}
    \caption{Comparison of the time-domain sample structures of (a) cyclic-prefix discrete multitone (CP-DMT) and (b) cyclic-prefix-free DMT (NoCP-DMT) carrying quadrature phase-shift keying (QPSK)-mapped payloads, where each CP-DMT block repeats \(N_{\rm cp}\) samples for an \(N\)-sample useful DMT symbol and the NoCP-DMT stream directly concatenates useful DMT symbols without cyclic redundancy.}
    \label{figCPvsNoCP}
\end{figure}

The full-memory CP enables low-complexity FDE but repeats \(N_{\mathrm{cp}}\) samples per useful symbol. Fig.~\ref{figCPvsNoCP} compares CP-DMT, for which \(\eta_{\mathrm{cp}}<1\), with NoCP-DMT, for which \(N_{\mathrm{cp}}=0\) and \(\eta_{\mathrm{cp}}=1\).

Let \(b\in\mathbb{Z}\) index the transmitted useful DMT symbols, independently of \(p\). Using the same transmit-derived reconstruction-reference definition as in Section~\ref{DMTFDE}, let \(q_b[n]\) denote the known reconstruction reference over the \(b\)-th useful DMT symbol. The NoCP-DMT effective excitation sequence on the DAS reconstruction grid is
\begin{equation}
    e_{\mathrm{NC}}[bN+n]
    =
    q_b[n],
    \label{eq:nocp_stream}
\end{equation}
where $0\le n\le N-1$, and the label \(\mathrm{NC}\) denotes the NoCP-DMT mode.
Successive useful DMT symbols are therefore directly concatenated without a copied prefix or a zero-valued intersymbol guard interval and can carry different payload data. On the forward link, eliminating the CP removes the full-memory sample overhead but does not by itself establish circular convolution. The equalizer is therefore determined by the memory of the end-to-end IM/DD channel; the receiver realization used in the experiment is described in Section~\ref{ExpDiss}. This forward-link equalization is independent of the backward DAS channel reconstruction developed below.

Without the CP condition in \eqref{eq:cp_condition}, the linear convolution in \eqref{eq:discrete_backscatter_isac} cannot be replaced by the blockwise circular model in \eqref{eq:dmt_circular_rx}. The contribution of samples transmitted before a DMT-symbol boundary must instead remain in the distributed sensing-channel model. For NoCP-DMT, \(e_p[n]\) in \eqref{eq:discrete_backscatter_isac} is the segment of \(e_{\mathrm{NC}}[n]\) aligned with the \(p\)-th observation interval. Retain \(N_{\mathrm{obs}}\) received samples in that interval and define
\begin{align}
    \mathbf y_p
    &\triangleq
    [y_p[0],\ldots,y_p[N_{\mathrm{obs}}-1]]^{\mathrm T},
   \label{eq:nocp_vectors11}\\
    \mathbf w_p
    &\triangleq
    [w_p[0],\ldots,w_p[N_{\mathrm{obs}}-1]]^{\mathrm T}.
    \label{eq:nocp_vectors}
\end{align}
For NoCP reconstruction, let \(M_{\mathrm{LS}}\ge M_{\min}\) be the selected LS support order, set \(M=M_{\mathrm{LS}}\) in \eqref{eq:discrete_backscatter_isac}, and zero-pad the physical channel when necessary. The known input context \(e_p[n]\), \(-(M_{\mathrm{LS}}-1)\le n\le N_{\mathrm{obs}}-1\), includes the \(M_{\mathrm{LS}}-1\) samples preceding the retained output window. Define \(\mathbf T_p\in\mathbb{C}^{N_{\mathrm{obs}}\times M_{\mathrm{LS}}}\) by
\begin{equation}
    [\mathbf T_p]_{r,m}
    =
    e_p[r-m],  
    \label{eq:nocp_toeplitz}
\end{equation}
where $0\le r\le N_{\mathrm{obs}}-1$ and $0\le m\le M_{\mathrm{LS}}-1$.
The finite-memory observation model is then given by
\begin{equation}
    \mathbf y_p
    =
    \mathbf T_p\mathbf h_p+\mathbf w_p.
    \label{eq:nocp_matrix_model}
\end{equation}
Accordingly, contributions spanning adjacent DMT symbols are not assumed to vanish; they are incorporated through the complete known input context in \(\mathbf T_p\).

The distributed sensing channel is reconstructed by regularized LS:
\begin{equation}
    \widehat{\mathbf h}^{\mathrm{LS}}_p
    =
    \arg\min_{\mathbf h}
    \left\|\mathbf y_p-\mathbf T_p\mathbf h\right\|_2^2
    +
    \lambda_{\mathrm{LS}}\left\|\mathbf h\right\|_2^2,
    \label{eq:nocp_regularized_ls}
\end{equation}
where \(\lambda_{\mathrm{LS}}\ge0\) is the regularization parameter. Consequently, the associated normal equation is
\begin{equation}
    \left(
    \mathbf T_p^{\mathrm H}\mathbf T_p
    +\lambda_{\mathrm{LS}}\mathbf I_{M_{\mathrm{LS}}}
    \right)
    \widehat{\mathbf h}^{\mathrm{LS}}_p
    =
    \mathbf T_p^{\mathrm H}\mathbf y_p,
    \label{eq:nocp_normal_equation}
\end{equation}
where \(\mathbf I_{M_{\mathrm{LS}}}\) is the \(M_{\mathrm{LS}}\times M_{\mathrm{LS}}\) identity matrix. When the coefficient matrix is nonsingular, the solution is
\begin{equation}
    \widehat{\mathbf h}^{\mathrm{LS}}_p
    =
    \left(
    \mathbf T_p^{\mathrm H}\mathbf T_p
    +\lambda_{\mathrm{LS}}\mathbf I_{M_{\mathrm{LS}}}
    \right)^{-1}
    \mathbf T_p^{\mathrm H}\mathbf y_p.
    \label{eq:nocp_ls_solution}
\end{equation}
Eq.~\eqref{eq:nocp_ls_solution} need not form or invert a dense Toeplitz matrix because products involving \(\mathbf T_p\) and \(\mathbf T_p^{\mathrm H}\) can be evaluated by linear convolution and correlation using the fast Fourier transform (FFT).

In the noiseless and unregularized case, unique recovery requires \(N_{\mathrm{obs}}\ge M_{\mathrm{LS}}\) and \(\operatorname{rank}(\mathbf T_p)=M_{\mathrm{LS}}\). Since \(M_{\mathrm{LS}}\ge M_{\min}\), full-memory reconstruction still requires an RTT-scale observation window. NoCP-DMT removes the transmitted CP, but not the physical memory or quasi-static requirement; positive \(\lambda_{\mathrm{LS}}\) improves stability at the cost of bias.

Let adjacent observation intervals be separated by \(N_{\mathrm{hop}}\) samples. The corresponding slow-time spacing and sampling rate are
\begin{equation}
    T_{\mathrm{hop}}=N_{\mathrm{hop}}T_s,
    \label{eq:nocp_slow_time}
\end{equation}
where \(f_{\mathrm{slow}}^{\mathrm{NC}}=1/T_{\mathrm{hop}}\). Under uniform slow-time sampling, an acoustic component at frequency \(f_{\mathrm{ac}}\) is nominally unaliased if \(f_{\mathrm{ac}}<f_{\mathrm{slow}}^{\mathrm{NC}}/2\). 

Unlike \(T_{\mathrm{PRI}}\) in pulsed LFM sensing, \(T_{\mathrm{hop}}\) is not constrained by the need to isolate successive pulse returns, because the effective sensing excitation remains continuous and the preceding channel memory is included in \eqref{eq:nocp_matrix_model}. The observation windows may therefore overlap. Nevertheless, the \(N_{\mathrm{obs}}\)-sample observation window, whose first-to-last sample span is \((N_{\mathrm{obs}}-1)T_s\), must remain within the quasi-static interval assumed in \eqref{eq:quasi_static_channel}. Its temporal aperture also shapes the usable acoustic response, and overlapping windows produce statistically correlated channel estimates.

The DFT of \(\widehat{\mathbf h}^{\mathrm{LS}}_p\) supplies \(\mathcal H_p[k]\) for the same PCC and gauge-differential phase recovery in \eqref{eq:pcc_channel}--\eqref{eq:gauge_phase}. This known-waveform Toeplitz estimation problem~\cite{MartinsOE2016} uses the transmitted NoCP-DMT sequence and thus removes the full-memory CP at the cost of regularized long-memory reconstruction.

\section{Experimental Results and Discussion}\label{ExpDiss}

\begin{figure}[!t]
    \centering
    \includegraphics[width=0.45\textwidth]{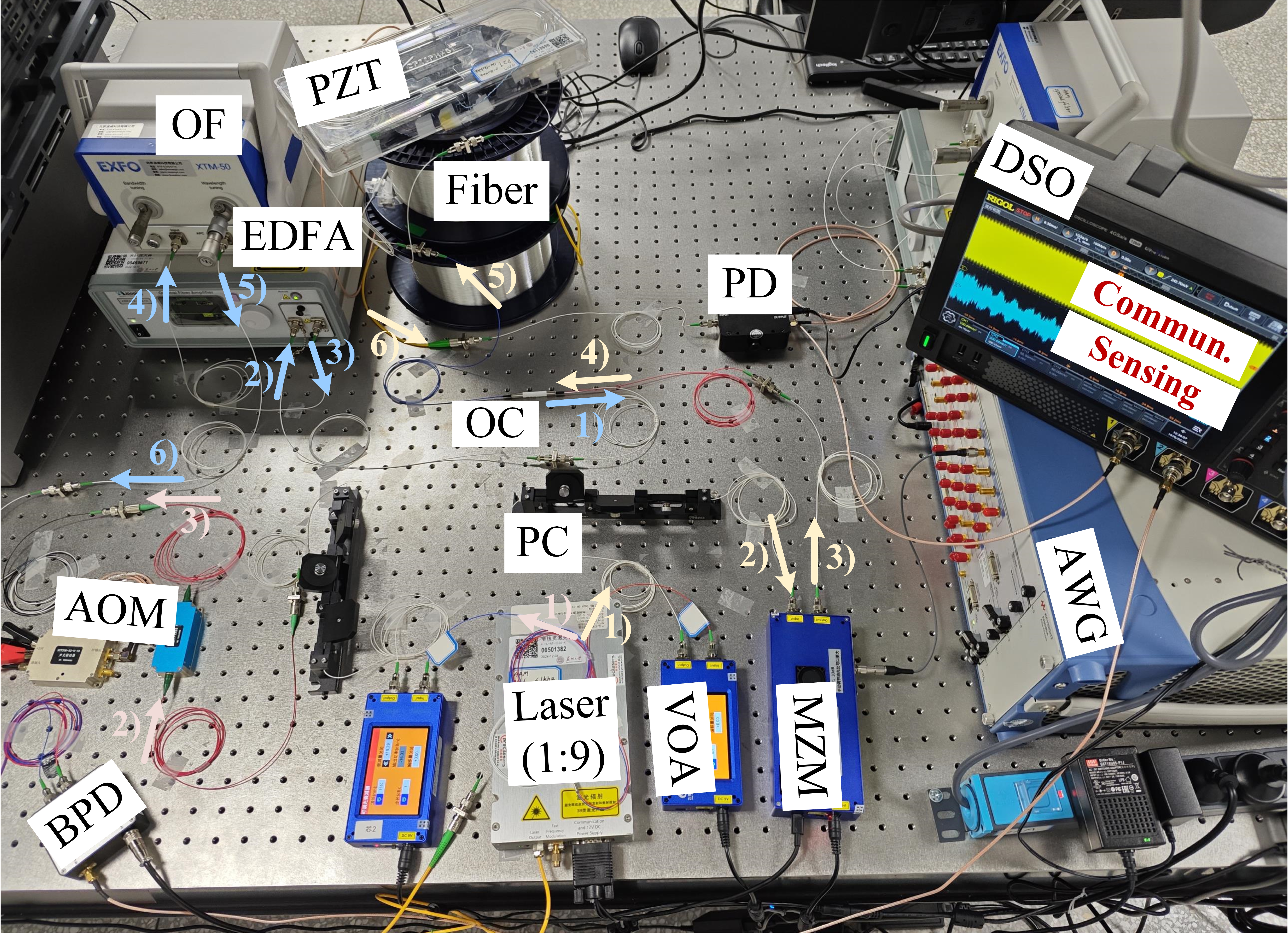}
    \caption{
    Experimental setup of the proposed continuous payload-bearing DMT framework for fiber-optic integrated sensing and communication (ISAC). AWG: arbitrary waveform generator; MZM: Mach--Zehnder modulator; VOA: variable optical attenuator; PC: polarization controller; OC: optical circulator; PD: photodetector; AOM: acousto-optic modulator; PZT: piezoelectric transducer; EDFA: erbium-doped fiber amplifier; OF: optical filter; BPD: balanced photodetector; DSO: digital storage oscilloscope.}
    \label{fig:Exp}
\end{figure}

The experimental setup used to validate the continuous DMT-based fiber-optic ISAC scheme is shown in Fig.~\ref{fig:Exp}. The output of a narrow-linewidth laser (FL-SF-1550-S, Precilasers, China) is divided by a 1:9 optical coupler into LO and signal branches. In the signal branch, a variable optical attenuator (VOA) and a polarization controller (PC) adjust the optical power and polarization, respectively. An arbitrary waveform generator (AWG; M8190A, Keysight Technologies, USA) generates either the continuous CP-DMT or NoCP-DMT waveform modes described in Section~\ref{DMTISAC}, and the selected electrical waveform then drives a 10-GHz Mach--Zehnder modulator (MZM; MZM-M-10-2.92F-FA, Fiber-Photonics, China) biased at the quadrature point. The same optical transmitter and fiber link are used for both waveform modes.

The modulated optical waveform is launched through an optical circulator (OC) into an approximately 10.1-km link comprising two approximately 5-km spools of standard single-mode fiber. A PZT (PZ1-SMF4-APC-E, Optiphase, Van Nuys, CA, USA) at the junction of the two spools emulates a localized vibration. At the remote end, the forward-propagating signal is directly detected by a 500-MHz InGaAs ultra-low-noise photodetector (PD; UPD-500M-A, LD-PD, Singapore), providing the IM/DD observation for payload recovery.

For DAS, the Rayleigh-backscattered field returns through the OC, is amplified by an erbium-doped fiber amplifier (EDFA; AEDFA-PA-30-B-FA, Amonics, Hong Kong, China), and is subsequently filtered by an optical filter (OF; XTM-50, EXFO, Canada) to suppress amplified spontaneous emission noise. In the LO branch, the 10\% split laser output passes through a PC and a 200-MHz frequency-shifting fiber-coupled AOM (SGTF200-1550-1, SMART SCI\&TECH, China). The 200-MHz-shifted LO and the filtered Rayleigh-backscattered field are coupled into a 500-MHz InGaAs ultra-low-noise BPD (UBD-500M-A, LD-PD, Singapore) for heterodyne coherent detection. The electrical outputs of the PD and BPD are synchronously acquired by two channels of a digital storage oscilloscope (DSO; DHO4404, RIGOL Technologies, China) at 1~GSa/s, with each record spanning approximately 100~ms.

For reconstruction, the DAS waveform and transmit-derived reference were placed on a common 400-MSa/s grid (\(T_s=2.5\)~ns); all sample counts below refer to this grid. The nominal 10.1-km link gives \(M_{\min}=40401\). CP-DMT used \(N=51200\), \(M=M_{\min}\), and \(N_{\mathrm{cp}}=41600\), corresponding to 128-, 104-, and 232-\(\mu\)s useful-symbol, CP, and block durations, respectively. NoCP-DMT concatenated \(N=3200\)-sample, 8-\(\mu\)s symbols and used \(M_{\mathrm{LS}}=41600\), \(N_{\mathrm{obs}}=76800\), and \(N_{\mathrm{hop}}=41600\). The NoCP LS order contains a 1199-tap (approximately 3-\(\mu\)s or 300-m) conservative margin beyond \(M_{\min}\); its observation and hop durations are 192 and 104~\(\mu\)s.

Both modes employed a two-sided DMT spectral aperture of 256~MHz, spanning \(-128\) to 128~MHz and corresponding to a one-sided electrical bandwidth of 128~MHz. After reserving center and outer-edge guard bands, the active subcarriers occupied \([-121.75,-3.125]\cup[3.125,121.75]\)~MHz. Because the real-valued DMT waveform was generated using Hermitian symmetry, the negative-frequency subcarriers were conjugate mirrors of their positive-frequency counterparts; CP-DMT and NoCP-DMT therefore contained \(15{,}185\) and 950 independent positive-frequency active subcarriers, respectively. The retained frequencies provided an outer-edge-to-edge reconstruction span of \(B_{\mathrm{rec}}=243.5\)~MHz, including the inactive \(\pm3.125\)-MHz center guard within the span. This reconstruction span gives a nominal full-band spatial resolution of approximately 0.41~m, whereas \(\Delta z=0.25\)~m is the sampling-grid interval. The FDE regularization was \(\lambda=10^{-3}P_{p,\mathrm{med}}\), where \(P_{p,\mathrm{med}}\triangleq\operatorname*{median}_{k\in\mathcal{K}_p}|Q_p[k]|^2\), and frequency bins outside the retained set were set to zero. For NoCP-DMT, \(\lambda_{\mathrm{LS}}=0.02\operatorname*{median}\{\operatorname{diag}(\mathbf T_p^{\mathrm H}\mathbf T_p)\}\). Both reconstructions used \(G=40\), corresponding to a 10-m gauge length.

\begin{figure}[!t]
    \centering
    \includegraphics[scale=0.081]{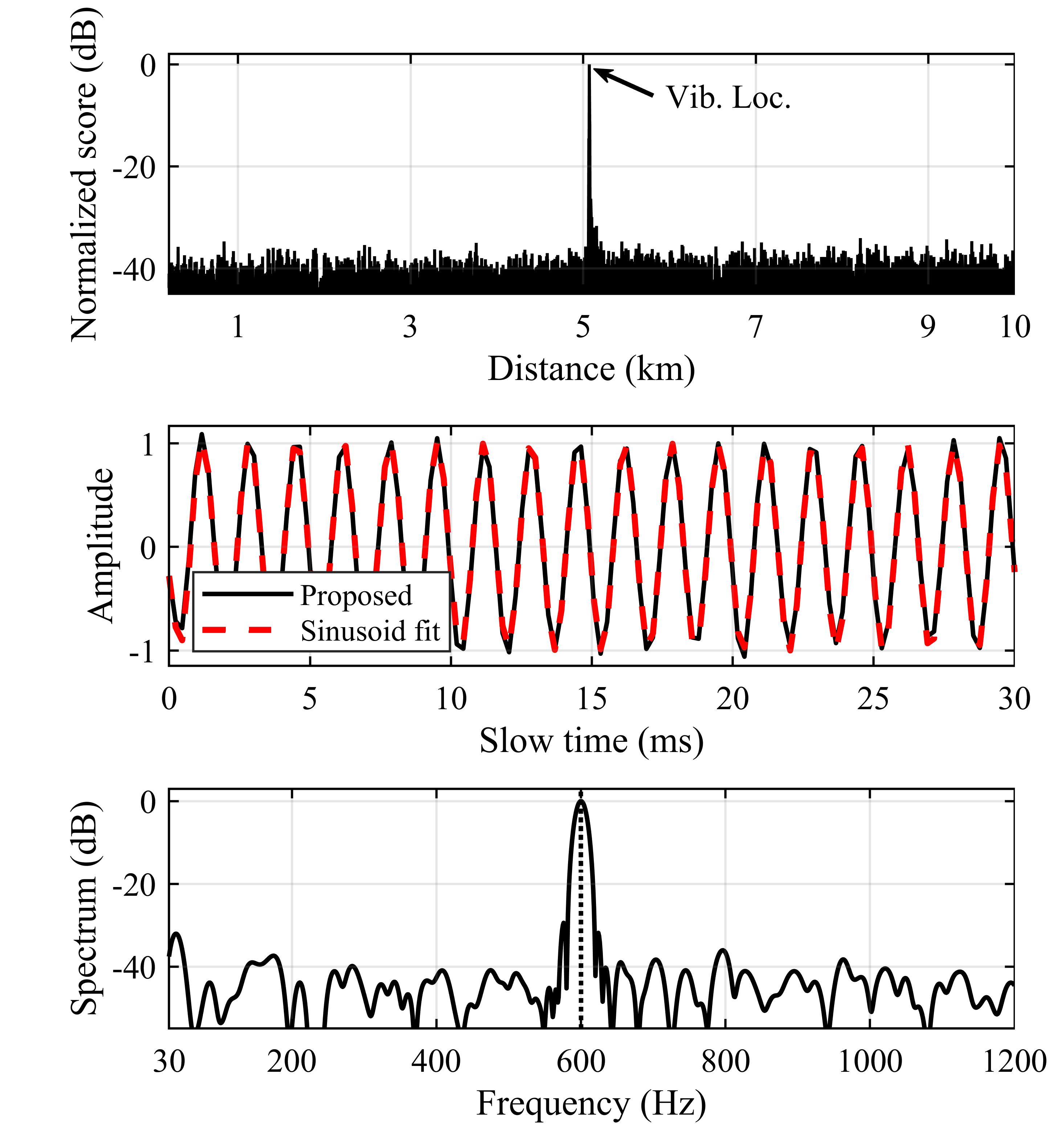}
    \caption{Experimental distributed acoustic sensing results for the continuous CP-DMT waveform under a 2-V, 600-Hz PZT drive, where the upper panel shows the peak-normalized single-tone projection score with its dominant peak at 5.07075~km, the middle panel compares the recovered gauge-differential phase trace with its sinusoidal fit, both normalized by the fitted amplitude, over a representative 30-ms portion of the approximately 100-ms record, with a full-record correlation coefficient of 0.9982, and the lower panel shows the peak-normalized phase PSD with the estimated frequency of 600.0125~Hz marked by a vertical line.}
    \label{ResfigCPDMT2V}
\end{figure}

\begin{figure}[!t]
    \centering
    \includegraphics[scale=0.081]{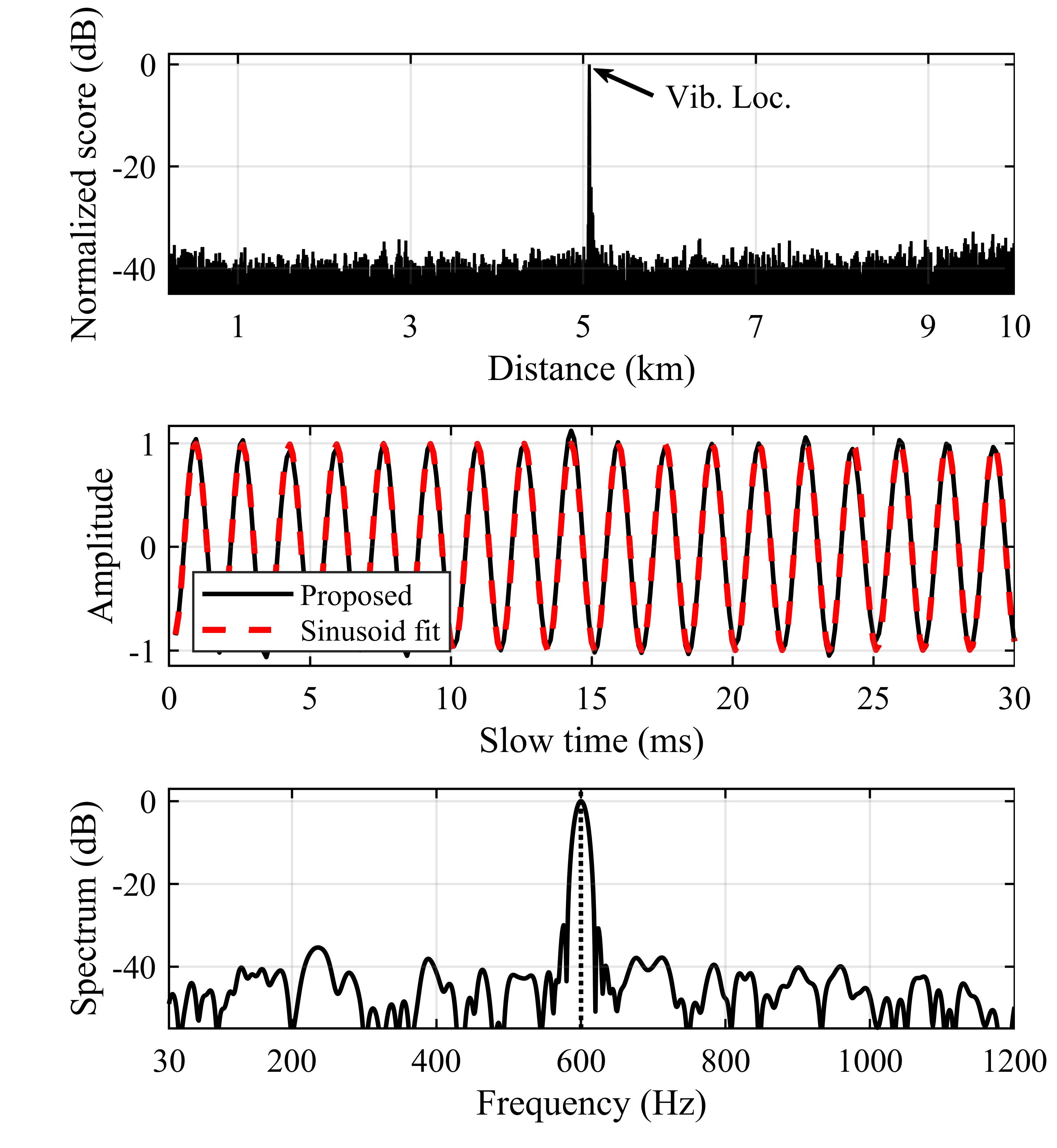}
    \caption{Experimental distributed acoustic sensing results for the continuous NoCP-DMT waveform under a 2-V, 600-Hz PZT drive, where the upper panel shows the peak-normalized single-tone projection score with its dominant peak at 5.07125~km, the middle panel compares the recovered gauge-differential phase trace with its sinusoidal fit, both normalized by the fitted amplitude, over a representative 30-ms portion of the approximately 100-ms record, with a full-record correlation coefficient of 0.9983, and the lower panel shows the peak-normalized phase PSD with the estimated frequency of 600.0099~Hz marked by a vertical line.}
    \label{ResfigNoCPDMT2V}
\end{figure}

For the DAS measurements shown in Figs.~\ref{ResfigCPDMT2V} and~\ref{ResfigNoCPDMT2V}, the PZT was driven by a 600-Hz sinusoidal voltage at a 2-V drive level. Following the respective distributed-channel reconstructions described in Sections~\ref{DMTFDE} and~\ref{NoCPDMT}, the same blind post-processing chain was applied to both sets of reconstructed channel responses. Both CP-DMT and NoCP-DMT carried QPSK-mapped payload data generated from an image bitstream, and their corresponding time-domain sample structures are illustrated in Fig.~\ref{figCPvsNoCP}. For forward communication, CP-DMT was recovered by pilot-aided one-tap FDE after CP removal. Complete NoCP-DMT symbols were recovered by pilot-aided per-subcarrier one-tap equalization, used here as a short-memory approximation for the tested forward link rather than as an exact CP-free circular-convolution model. %The reported EVM and bit-error results therefore include any residual intercarrier and intersymbol interference under the tested link conditions.

The reconstructed responses were processed by PCC and gauge-differential phase recovery in \eqref{eq:pcc_channel}--\eqref{eq:gauge_phase}. The whitening model propagated white receiver noise through the FDE inverse and spectral windows for CP-DMT; for NoCP-DMT, it used a spectral-window Gram approximation assuming white channel-spectrum noise. Combining weights and endpoint reference amplitudes were calibrated on an initial portion of each record and then fixed. %Neither the PZT position nor its drive frequency was supplied to the blind search.
For each retained gauge, linear detrending, zero-phase bandpass filtering over \(\mathcal B=[f_{\min},f_{\max}]=[30,1200]\)~Hz, and DC removal were applied independently to reliable continuous phase intervals. Let \(\psi_\ell[p]\) denote the processed phase. Its Hann-windowed power spectral density (PSD), evaluated at the measured slow-time timestamps, was divided by the spatial median at each frequency to identify tone candidates within \(\mathcal B\). Candidate frequencies were refined by sinusoidal fitting, with the sine and cosine templates subjected to the same preprocessing as the measured phase.

At the selected candidate frequency \(f_\star\), let \(\widehat\psi_\ell[p;f_\star]\) denote the least-squares fit using the preprocessed sine and cosine templates with one common coefficient pair over the retained full-record observations. The projected-energy fraction \(R_\ell^2(f_\star)\) and single-tone projection score are
\begin{equation}
    R_\ell^2(f_\star) =\frac{\displaystyle\sum_{p\in\mathcal O_\ell}
        |\widehat\psi_\ell[p;f_\star]|^2}
    {\displaystyle\sum_{p\in\mathcal O_\ell}|\psi_\ell[p]|^2}, 
    \label{eq:exp_localization_score}
\end{equation}
and 
\begin{equation}
    S_\ell(f_\star) =10\log_{10}
    \frac{\max\{R_\ell^2(f_\star),\epsilon_{\min}\}}
    {\max\{1-R_\ell^2(f_\star),\epsilon\}},
    \label{eq:exp_localization_score1}
\end{equation}
where \(\mathcal O_\ell\) contains the jointly retained phase and template observations over the complete record, and \(\epsilon_{\min}>0\) and \(\epsilon>0\) are numerical floors. Maximizing \(S_\ell(f_\star)\) over reliable gauge positions determines the location, followed by frequency refinement at that position. The plotted localization profile is \(S_\ell(f_\star)-\max_j S_j(f_\star)\), with the maximum taken over reliable gauge positions. The recovered phase is compared with its fitted sinusoid using the absolute Pearson correlation coefficient over the retained full-record observations.

Fig.~\ref{ResfigCPDMT2V} presents the result obtained after the one-tap sensing-channel FDE of the CP-DMT record. The single-tone projection score exhibits a unique dominant peak at 5.07075~km, in the PZT-containing junction region of the two fiber spools. The recovered gauge-differential phase trace follows its sinusoidal fit over the retained full-record observations with a correlation coefficient of 0.9982, and the estimated vibration frequency is 600.0125~Hz. The spatial, temporal, and spectral results show recovery of the vibration-induced phase evolution.

Fig.~\ref{ResfigNoCPDMT2V} applies the same evaluation to the NoCP-DMT record after regularized LS channel reconstruction. The single-tone projection score has its largest peak at 5.07125~km. The recovered gauge-differential phase trace achieves a correlation coefficient of 0.9983 with its sinusoidal fit over the retained full-record observations, and the estimated vibration frequency is 600.0099~Hz. Thus, the NoCP-DMT mode retains blind vibration localization and phase-trace recovery while eliminating the full-memory CP, at the cost of the long-memory channel-reconstruction complexity.

Taken together, the two records identify the PZT-containing central section of the approximately 10-km link and recover its 600-Hz sinusoidal excitation using the same PCC-based phase-recovery procedure. These results support both receiver realizations of the continuous payload-bearing DMT framework: CP-DMT provides low-complexity one-tap sensing-channel FDE through cyclic redundancy, whereas NoCP-DMT exchanges that redundancy for regularized LS reconstruction of the linear-convolution channel.

\begin{figure}[!t]
    \centering
    \includegraphics[scale=0.085]{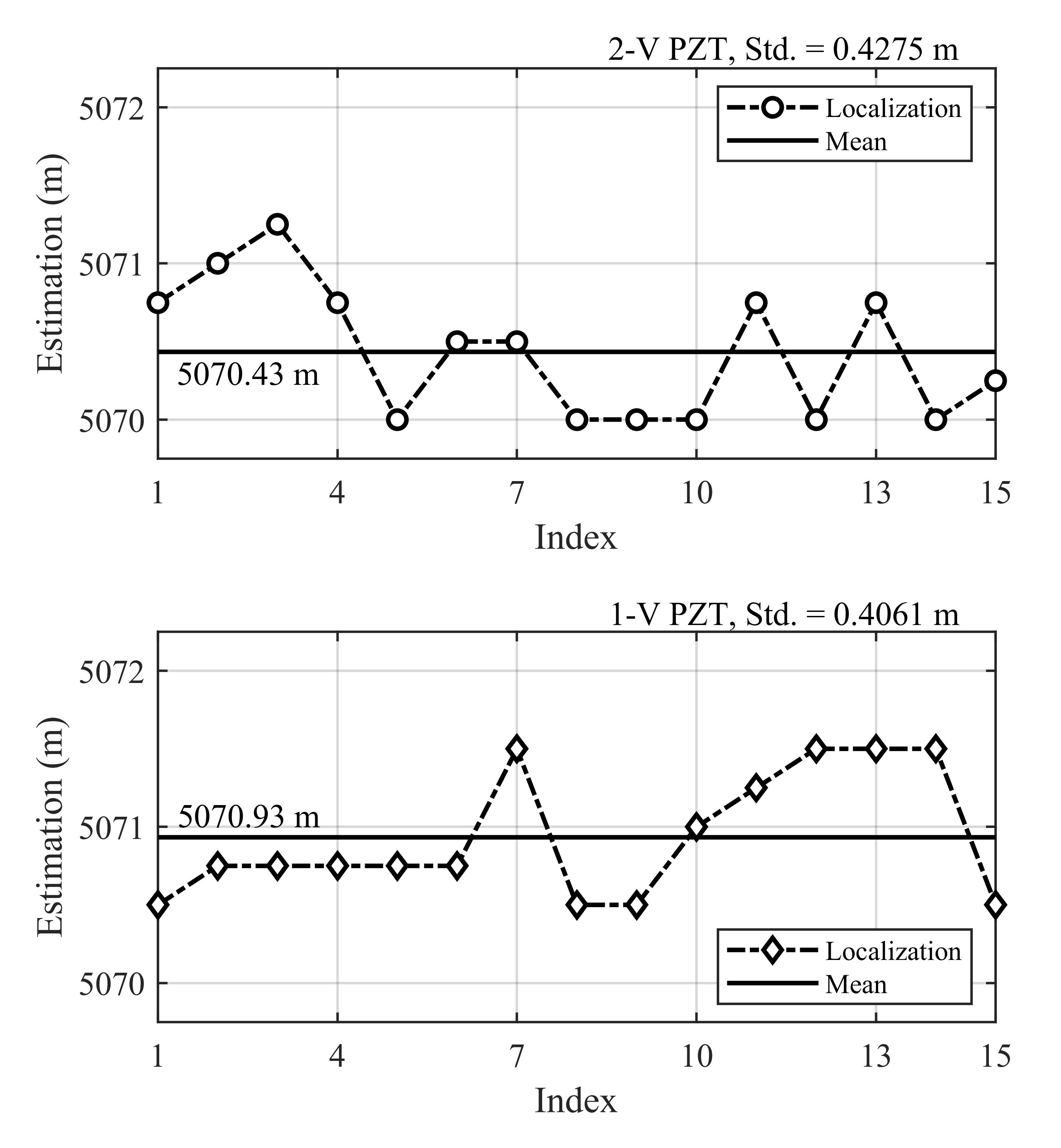}
    \caption{CP-DMT sensing results for fifteen selected records per drive level under 2-V (upper) and 1-V (lower), 600-Hz PZT drives, where the 2- and 1-V results have mean positions of 5070.4333 and 5070.9333~m and standard deviation (Std.) of 0.4275 and 0.4061~m, respectively.}
    \label{ResfigCPDMT2VStd}
\end{figure}

\begin{figure}[!t]
    \centering
    \includegraphics[scale=0.085]{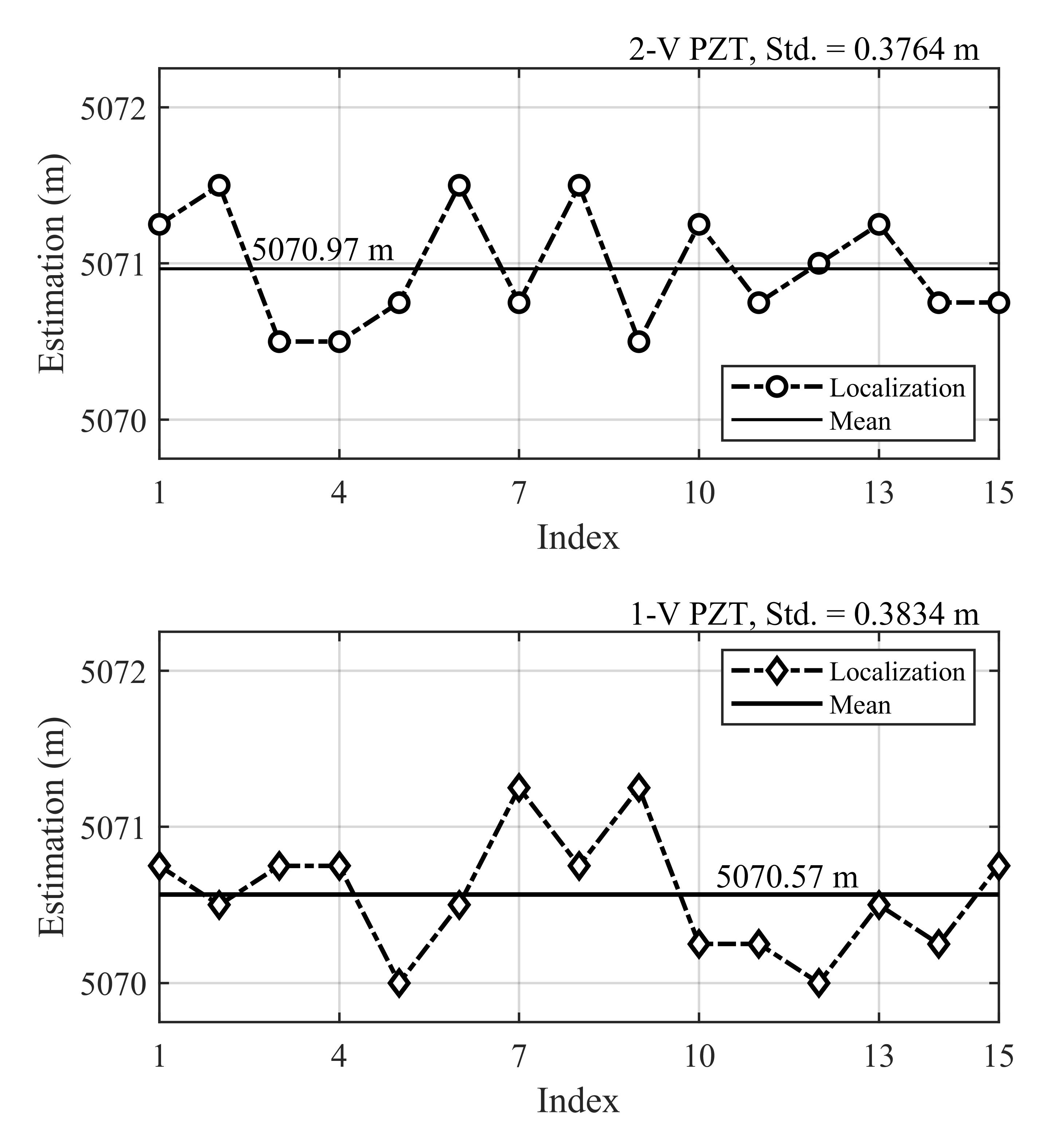}
    \caption{NoCP-DMT sensing results for fifteen selected records per drive level under 2-V (upper) and 1-V (lower), 600-Hz PZT drives, where the 2- and 1-V results have mean positions of 5070.9667 and 5070.5667~m and standard deviation (Std.) of 0.3764 and 0.3834~m, respectively.}
    \label{ResfigNoCPDMT2VStd}
\end{figure}

To examine record-to-record localization variation, fifteen DAS records are selected were analysis for each waveform at each of the 2- and 1-V PZT drive levels. The corresponding localization results for CP-DMT and NoCP-DMT are illustrated in Figs.~\ref{ResfigCPDMT2VStd} and~\ref{ResfigNoCPDMT2VStd}, respectively. As shown in Fig.~\ref{ResfigCPDMT2VStd}, the CP-DMT localization estimates span 5070.00--5071.25~m at 2~V and 5070.50--5071.50~m at 1~V. The sample standard deviation at 2~V is 0.4275~m, slightly larger than the 0.4061~m at 1~V. %The difference of 0.0214~m is smaller than the localization-grid spacing of \(\Delta z=0.25\)~m obtained from~\eqref{eq:delta_z} for the \(F_s=400\)~MSa/s reconstruction grid.
For NoCP-DMT, Fig.~\ref{ResfigNoCPDMT2VStd} shows sample standard deviations of 0.3764~m at 2~V and 0.3834~m at 1~V. The corresponding localization estimates span 5070.50--5071.50~m and 5070.00--5071.25~m, respectively, and both drive levels consistently identify the same PZT-containing junction region.

%Accordingly, this sub-grid difference should not be interpreted as evidence of higher localization precision at the lower drive level. With fifteen selected records, such a small difference can be influenced by range-grid quantization and record-to-record variations in the recovered sensing response. 

The interchannel-delay anchor accounts for the nominal one-way fiber delay but is not externally calibrated to remove residual hardware skew. Accordingly, the reported positions should be interpreted as internally consistent, rather than externally calibrated, absolute-range coordinates. The small differences between the CP-DMT and NoCP-DMT mean positions should therefore not be interpreted as physical displacement of the PZT. Because each condition uses fifteen records selected, the reported values are insufficient to establish a statistically resolved dependence on the PZT drive voltage. 
A larger number of repeated acquisitions would be required for a rigorous statistical characterization of localization precision.

\begin{table*}[!t]
    \caption{Summary of the two continuous payload-bearing DMT modes for fiber-optic ISAC performance.}
    \label{tab:ExpJointSummary}
    \centering
    \renewcommand{\arraystretch}{1.15}
    \setlength{\tabcolsep}{9pt}
    \begin{tabular}{|l|c|c|c|c|}
        \hline
        Scheme & \multicolumn{2}{c|}{CP-DMT} & \multicolumn{2}{c|}{NoCP-DMT} \\
        \hline
        PZT drive & 2~V & 1~V & 2~V & 1~V \\
        \hline
        Localization Std./mean (m) & 0.43/5070.43 & 0.41/5070.93 & 0.38/5070.97 & 0.38/5070.57 \\
        \hline
        Mean IM/DD EVM (\%) & 4.86 & 4.79 & 4.95 & 5.01 \\
        \hline
    \end{tabular}
\end{table*}

The CP-DMT and NoCP-DMT modes are evaluated under their respective waveform and reconstruction parameterizations. The following results demonstrate the feasibility of both operating modes rather than an all-else-equal performance ranking between CP-DMT and NoCP-DMT.
Table~\ref{tab:ExpJointSummary} summarizes the sensing and communication results for the same fifteen selected records per condition. Mean EVM ranges from 4.79\% to 5.01\% with no bit errors occurred, and all recovered images passed the cyclic redundancy check. Therefore, both modes maintain stable forward payload recovery while supporting backward DAS under the tested PZT drives.
 
\section{Conclusions}\label{Conclu}
This paper developed and experimentally demonstrated a novel fiber-optic ISAC framework with two continuous payload-bearing DMT modes. In either mode, the same waveform carries the forward payload and excites backward DAS. The finite-memory sensing model distinguishes the RTT-scale return-isolation requirement from MF-correlation-induced spatial ISI. CP-DMT uses sufficient CP and FDE to reconstruct grid-aligned taps without this MF-induced coupling under ideal full-bin inversion, while NoCP-DMT removes the CP through regularized LS at higher receiver complexity. Finite reconstruction bandwidth and regularization can still cause practical spreading.

Over an approximately 10-km link, both modes blindly localized the 600-Hz PZT disturbance near the internally referenced 5.071-km coordinate; the representative 2-V records yielded gauge-differential phase correlations of 0.9982 and 0.9983 for CP-DMT and NoCP-DMT, respectively. Across the four fifteen-record conditions, the localization standard deviations were 0.38--0.43~m and the mean IM/DD EVM values were 4.79\%--5.01\%, with no bit errors. These finite samples demonstrate record-to-record repeatability but do not establish a drive-voltage dependence. The results establish the feasibility of common-waveform fiber-optic ISAC, in which the payload-bearing communication waveform itself serves as the sensing excitation, and provide a foundation for access links that combine data transmission with distributed monitoring without a separate sensing waveform.

\end{document}